\documentclass[article,shortnames]{jss}%
\usepackage{thumbpdf,lmodern}

\usepackage{comment}
\usepackage{amssymb}
\usepackage{bm}
\usepackage{amsmath}
\usepackage{booktabs}
\usepackage{comment}
\usepackage{subfigure}
\usepackage{framed}
\usepackage{amsfonts}
\usepackage{graphicx}%
\usepackage{hyperref}
\graphicspath{{}}

\author{Yaqiong Wang \\ Peking University \And Francesco Finazzi\\ University of Bergamo \And Alessandro Fass\`{o} \\ University of Bergamo}
\Plainauthor{Yaqiong Wang, Francesco Finazzi, and Alessandro Fass\`{o}}

\title{\pkg{D-STEM} v2: A Software for Modelling Functional Spatio-Temporal Data}
\Plaintitle{Modelling functional spatio-temporal data with D-STEM}
\Shorttitle{Modelling functional spatio-temporal data with D-STEM}

\Abstract{Functional spatio-temporal data naturally arise in many environmental and climate applications where
data are collected in a three-dimensional space over time. 
The \proglang{MATLAB} \pkg{D-STEM} v1 software package was first introduced for modelling multivariate space-time data and has been recently extended to \pkg{D-STEM} v2 to handle functional data indexed across space and over time.
This paper introduces the new modelling capabilities of \pkg{D-STEM} v2 as well as the complexity reduction techniques required when dealing with large data sets. Model estimation, validation
and dynamic kriging are demonstrated in two case studies, one related to ground-level air quality data in Beijing, China, and the other one related to atmospheric profile data collected globally through radio sounding.}

\Plainkeywords{functional data analysis, 4D data, climate data, environmetrics, EM algorithm}

\Address{
Yaqiong Wang\\
Guanghua School of Management\\
Peking University\\
Yiheyuan road, 5\\
100871 Peking, China\\
E-mail: \email{yaqiongwang@pku.edu.cn}\\

Francesco Finazzi\\
Department of Management, Information and Production Engineering\\
University of Bergamo\\
viale Marconi, 5\\
24044 Dalmine (BG), Italy\\
E-mail: \email{francesco.finazzi@unibg.it}\\
URL: \url{http://www.unibg.it/pers/?francesco.finazzi}\\

Alessandro Fass{\`o}\\
Department of Management, Information and Production Engineering\\
University of Bergamo\\
viale Marconi, 5\\
24044 Dalmine (BG), Italy\\
E-mail: \email{alessandro.fasso@unibg.it}\\
URL: \url{http://www.unibg.it/pers/?alessandro.fasso}\\
}

\begin{document}
\section[Introduction]{Introduction}\label{sec:intro}

With the increase of multidimensional data availability and modern computing power, statistical models for spatial 
and spatio-temporal data are developing at a rapid pace. Hence, there is a need for stable and reliable, yet updated and
efficient, software packages. In this section, we briefly discuss multidimensional data in climate and environmental studies as well as statistical software for space-time data. 

\subsection{Multidimensional data}\label{sec:multidata}

Large multidimensional data sets often arise when climate and environmental phenomena are observed at the global
scale over extended periods. In climate studies, relevant physical variables are observed on a three-dimensional (3D) 
spherical shell (the atmosphere) while time is the fourth dimension. 
For instance, measurements are obtained by radiosondes flying from ground level up to the stratosphere \citep{fasso2014statistical}, by interferometric sensors aboard
satellites \citep{finazzi2018statistical} or by laser-based methods, such as Light Detection and Ranging (LIDAR) \citep{negri2018modeling}. 
In this context, statistical modelling of multidimensional data requires describing and exploiting the spatio-temporal correlation of the underlying phenomenon or data-generating process. 
This is done using explanatory variables and multidimensional latent variables with covariance functions defined over a convenient spatio-temporal support. 
When considering 3D$\times$T data (4D for brevity), covariance functions defined over the 4D support may be adopted.
However, these covariance functions often have a complex form \citep{Porcu2018}.
Moreover, when estimating the model parameters or making inferences,
very large covariance matrices (though they may be sparse) are implied.\\
In large climate and environmental applications, 4D data are rarely collected at high frequency in all spatial and temporal dimensions. 
Often, only one dimension is sampled at high frequency while the remaining dimensions are sampled sparsely. 
Radiosonde data, for instance, are sparse over the Earth's sphere, but they are dense along the vertical dimension, providing atmospheric profiles. 
This suggests that handling all spatial dimensions equally (e.g. using a 3D covariance function) may not be the best option from a modelling or computational perspective, 
and a data reduction technique may be useful instead. In this paper, the functional data analysis (FDA) approach 
\citep{ramsay2007applied} is adopted to model the relationship between measurements along the profile, while the remaining dimensions 
are handled following the classic spatio-temporal data modelling approach using only 2D spatial covariance functions.

\subsection{Statistical software}

Various software programmes are available for considering data on a plane or in a two-dimensional (2D) Euclidean space. 
The choice is more restricted when considering multidimensional or non-Euclidean spaces arising from atmospheric or remote sensing spatio-temporal data observed
on the surface of a sphere and over time.

\begin{figure}
\centering
\includegraphics[width=0.9\textwidth]{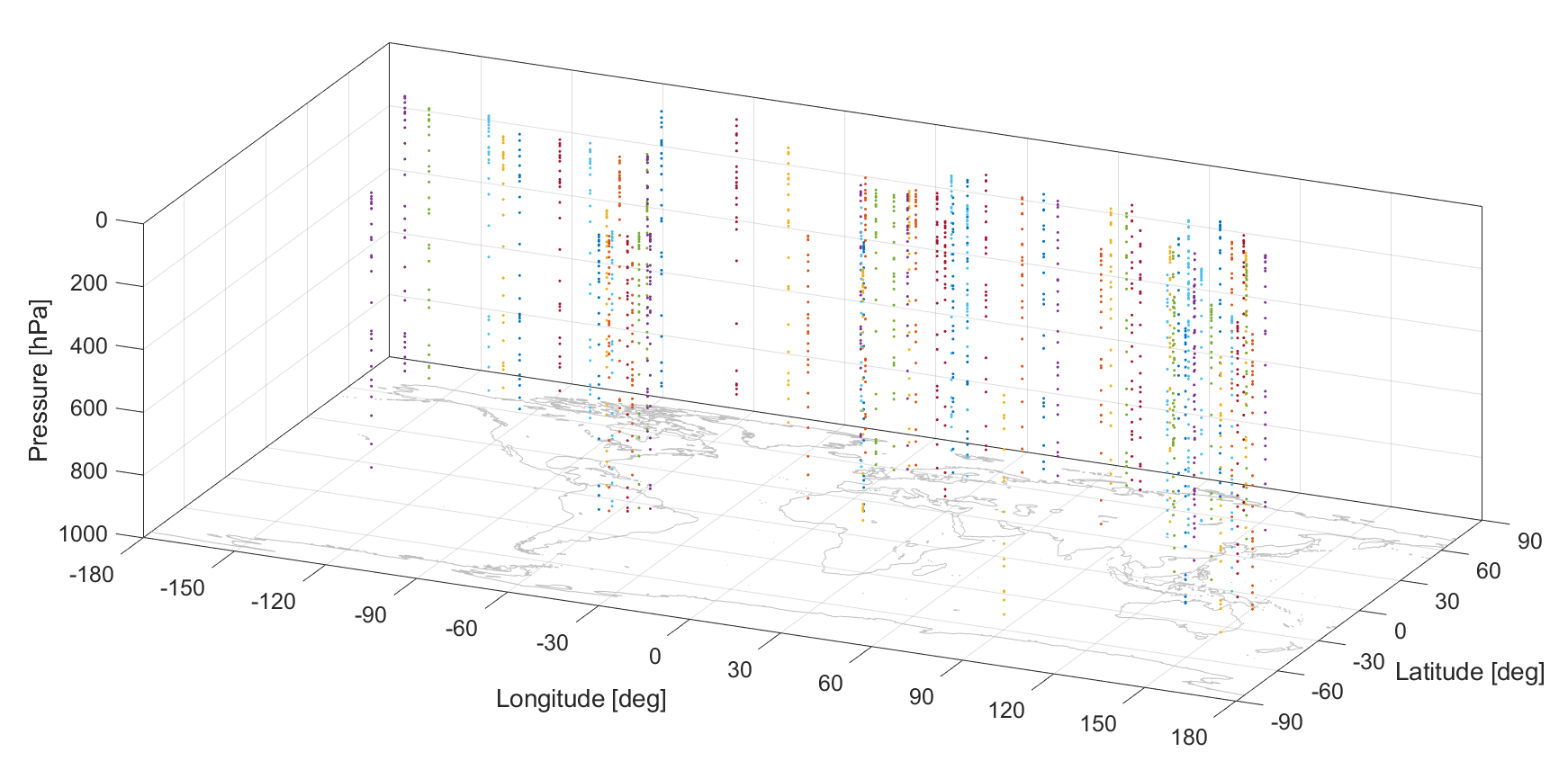}
\caption{Radio sounding data example. Each dot represents the spatial location of a measurement taken by a radiosonde. Dots of the same colour belong to the same radiosonde. (Pressure axis not in scale).}
\label{fig:RAOB_data}
\end{figure}

For example, Figure \ref{fig:RAOB_data} depicts
the spatial locations of measurements collected globally in a single day through radio sounding, as discussed in Section \ref{sec:casestudy_climate}. 
Space is three-dimensional, and measurements are repeated over time at the same spatial locations over the Earth's surface but at different pressure values.

The \pkg{spBayes} package \citep{finley2015} handles large spatio-temporal data sets, but space is only 2D. The documentation of the \pkg{spacetime} \citep{pebesma2012} and \pkg{gstat} \citep{pebesma2016} packages does not explicitly address the multidimensional case, but, according to \cite{gasch2015}, both packages have some capabilities to handle the 3D$\times$T.
However, we want to avoid working with 3D spatial covariance functions or sample spatio-temporal variograms. 
Fixed rank kriging (\citeauthor{cressie2008fixed} \citeyear{cressie2008fixed}) implemented in the \proglang{R} 
package \pkg{FRK} \citep{zammit2018frk} handles spatial and  spatio-temporal data both on the Euclidean plane and on the 
surface of the sphere. \pkg{FRK} implements a set of tools for data gridding and basis function computation, resulting in 
efficient dimension reduction, allowing it to handle large satellite data sets \citep{cressie2018mission}. 
It is based on a spatio-temporal random effects (SRE) model estimated by the 
expectation-maximisation (EM) algorithm. 
Recent extensions to \pkg{FRK} include the use of multi-resolution basis functions \citep{tzeng2018resolution}.

A second package based on SRE and the EM algorithm is \pkg{D-STEM} v1 \citep{finazzi2014d}. 
This package implements an efficient state-space approach for handling the temporal dimension and a heterotopic multivariate response approach that is useful when correlating heterogeneous networks \citep{fasso2011,calculli2015maximum}.

\pkg{D-STEM} v1 has been successfully used in various medium-to-large applications, proving that the EM algorithm implementation, being mainly based on closed-form iterations, is quite stable.
These applications include air quality assessment in the metropolitan areas of Milan, Teheran and Beijing \citep{fasso2013,Taghavi2019,Wan2020ENV};
multivariate spatio-temporal modelling at the country and continental levels in Europe \citep{finazzi2013,fasso2016};
time series clustering \citep{finazzi2015}; 
emulators of atmospheric dispersion modelling systems \citep{finazzi2019}; and 
near real-time prediction of earthquake parameters \citep{finazzi2020}.\\

A brief, non-exhaustive list of other models and/or software packages for advanced spatial data modelling is presented below, according to the principal technique, allowing the handling of large data sets. 
In general, these techniques aim at avoiding the Cholesky decomposition of large and dense covariance matrices.

Some approaches, including \pkg{FRK} and \pkg{D-STEM} v1, leverage sparse variance--covariance matrices. Others exploit the sparsity of the precision matrix, thanks to a spatial Markovian assumption. This class includes the \proglang{R} packages \pkg{LatticeKrig} \citep{nychka2015multiresolution,Nychka2016}, \pkg{INLA}
\citep{blangiardo2013,lindgren2015,Bivand2015,rue2014} and the multi-resolution approximation approach of \cite{katzfuss2017multi}, 
which uses the predictive process and the state space representation \citep{jurek2018multi} to model spatio-temporal data. Low-rank models are another
popular approach used by \pkg{spBayes}. Finally, the \proglang{R} package \pkg{laGP} \citep{gramacy2016lagp}, based on a machine learning approach, implements an efficient nearest neighbour prediction-oriented method. 
\cite{heaton2018case} develop an interesting spatial prediction competition considering a large data set and involving the above-mentioned approaches.

We observe that, although some of the software packages mentioned above consider both space and time, to the best of our knowledge, none of them handles a spatio-temporal FDA approach for data sets of the kind discussed in \ref{sec:multidata}.

In this paper, we present \pkg{D-STEM} v2, which is a \proglang{MATLAB} package, extending \pkg{D-STEM} v1. The new version introduces modelling of functional data indexed over space and time. 
Moreover, new complexity reduction techniques have been added for both model estimation and dynamic mapping, which are especially useful for large data sets. \\
The rest of the paper is organised as follows. 
Section \ref{sec:methodology} introduces the methodology adopted in this paper and, in particular, the data modelling approach and the complexity-reduction techniques. 
Section \ref{sec:software} describes the \pkg{D-STEM} v2 software in terms of the \proglang{MATLAB} classes used to define the data structure, model fitting and diagnostics and kriging. 
This is followed by an illustration of the software use through two case studies. 
The first one, discussed in Section \ref{sec:casestudy_ozone}, considers high-frequency spatio-temporal ozone data in Beijing. 
The second one, in Section \ref{sec:casestudy_climate}, considers modelling of global atmospheric temperature profiles and exploits the complexity-reduction capabilities of the new package. Finally, concluding remarks are provided in Section \ref{sec:remarks}.

\section[Methodology]{Methodology}\label{sec:methodology}

This section discusses the methodology behind the modelling and the complexity-reduction techniques implemented in \pkg{D-STEM} v2 
when dealing with functional space-time data sets. Moreover, model estimation, validation and dynamic kriging are briefly discussed.

\subsection{Model equations}

Let $\bm{s}=(s_{lat},s_{lon})^{ \top }$ be a generic spatial location on the
Earth's sphere, $\mathbb{S}^{2}$, and $t\in \mathbb{N} $ a discrete time index. 
It is assumed that the function of interest, $f\left(  \bm{s},h,t\right )$, with domain $\mathcal{H}=\left[  h_{1},h_{2}\right]  \subset \mathbb{R}$, can be observed at any $\left(  \bm{s},t\right)  $ and $h\in\mathcal{H}$ through noisy measurements $y(\bm{s},h,t)$ according to the following model:%
\begin{align}
y(\bm{s},h,t)  &  =f\left(  \bm{s},h,t\right)  +\varepsilon
(\bm{s},h,t),\label{eq:model_line1}\\
f\left(  \bm{s},h,t\right)   &  =\bm{x}(\bm{s},h,t)^{ \top }\bm{\beta}\left(  h\right)  +\bm{\phi}(h)^{\top}\bm{z}(\bm{s}%
,t),\label{eq:model_line2}\\
\bm{z}(\bm{s},t)  &  =\bm{G}\bm{z}(\bm{s}%
,t-1)+\bm{\eta}(\bm{s},t). \label{eq:model_line3}%
\end{align}
This model is referred to as the functional hidden dynamic geostatistical model (f-HDGM). In Equation (\ref{eq:model_line1}), $\varepsilon$ is a zero-mean Gaussian
measurement error independent in space and time with functional variance
$\sigma_{\varepsilon}^{2}\left(  h\right)  $, implying
that $\varepsilon$ is heteroskedastic across the domain $\mathcal{H}$. The variance is modelled as
\[
\log(\sigma_{\varepsilon}^{2}\left(  h\right)  )=\bm{\phi}(h)^{\top
}\bm{c}_{\varepsilon},
\]
where $\bm{\phi}(h)$ is a $p\times1$ vector of basis functions evaluated at
$h$, while $\bm{c}_{\varepsilon}$ is a vector of coefficients to be estimated. In Equation (\ref{eq:model_line2}), $\bm{x}(\bm{s},h,t)$ is a $b \times 1$ vector of covariates while $\bm{\beta}\left(  h\right)  =(\beta_{1}(h),...,\beta_{b}(h))^{\top}$ is the vector of functional parameters modelled as
\[
\beta_{j}(h)=\bm{\phi}(h)^{\top}\bm{c}_{\beta,j}, j=1,...,b,
\]
and $\bm{c}_{\beta}=\left( \bm{c}_{\beta,1}^{\top},...,\bm{c}_{\beta,b}^{\top}\right) ^{\top}$ is a $pb\times1$  vector of coefficients that needs to be estimated. Additionally, $\bm{z}(\bm{s},t)$ is a $p\times1$ latent space-time
variable with Markovian dynamics given in Equation (\ref{eq:model_line3}).
The matrix $\bm{G}$ is a diagonal transition matrix with diagonal elements in the $p \times 1$ vector
$\bm{g}$. The innovation vector $\bm{\eta}$ is obtained from a
multivariate Gaussian process that is independent in time but correlated across space
with matrix spatial covariance function given by
\[
\bm{\Gamma}(\bm{s},\bm{s}^{\prime};\bm{\theta})=diag\left(
v_{1}\rho(\bm{s},\bm{s}^{\prime};\bm{\theta}_{1}),...,v_{p}%
\rho(\bm{s},\bm{s}^{\prime};\bm{\theta}_{p})\right),
\]
where $\bm{v}=\left(  v_{1},...,v_{p}\right)  ^{\top}$ is a vector of variances and $\rho(\bm{s},\bm{s}^{\prime};\bm{\theta}_{j})$ is a valid spatial correlation function for locations $\bm{s},\bm{s}^{\prime}\in\mathbb{S}^{2}$, parametrised by $\bm{\theta}_{j}$, and $\bm{\theta}=(\bm{\theta}_{1},...,\bm{\theta}_{p})^{\top}$. The unknown model parameter vector is given by $\bm{\psi}=\left(
\bm{c}_{\varepsilon}^{\top},\bm{c}_{\beta}^{\top},\bm{g}^{\top},\bm{v}^{\top},\bm{\theta}^{\top}\right) ^{\top} $.

Note that, in order to ease the notation, the same $p$-dimensional basis functions $\bm{\phi}(h)$ are used to model
$\sigma_{\varepsilon}^{2}$, $\beta_{j}$ and $\bm{\phi}(h)^{\top}\bm{z}(\bm{s},t)$ in Equations 
(\ref{eq:model_line1})-(\ref{eq:model_line3}). In practice, \pkg{D-STEM} v2 allows one to specify a different number 
of basis functions for each model component. Also note that $\varepsilon$ is not a pure measurement error since it 
also accounts for model misspecification. 
Finally, the covariates $\bm{x}(\bm{s},h,t)$ are assumed to be known without error for any $\bm{s}$, $h$ and $t$, 
and thus they do not need a basis function representation.

\subsection{Basis function choice}

Choosing basis functions essentially means choosing the basis type and the number of basis functions. \pkg{D-STEM} v2 currently supports Fourier bases and B-spline bases. The former guarantee that the function is periodic in the domain $\mathcal{H}$, while the latter are not (in general) periodic
but have higher flexibility in describing functions with a complex shape.
Whichever basis function type is adopted, the number $p$ of basis functions must be fixed before model estimation. Usually, a high $p$ implies 
a better model $R^2$, but over-fitting may be an issue. Moreover, special care must be taken when choosing the number of basis functions for 
$\bm{\phi}(h)^{\top}\bm{z}(\bm{s},t)$. The classic FDA approach suggests fixing a high number of basis functions and adopting penalisation to avoid over-fitting. 
In our context, this is not viable since the covariance matrices involved in model estimation have dimension $n^3p^3 \times n^3p^3$. Since $n$ is usually large, a large $p$ would make model estimation unfeasible, especially if the number of time points $T$ is also high. 
When using B-spline basis, a small $p$ implies that the location of knots along the domain $\mathcal{H}$ also matters and may affect the model fitting performance. Ideally, $p$ and knot locations are chosen using a model validation technique (see \ref{sec:val}) by trying different combinations of $p$ and knot locations. 
If, due to time constraints, this is not possible, equally spaced knots are a convenient option.

\subsection{Model estimation}\label{sec:model_estimation}

The estimation of $\bm{\psi}$ and the latent space-time variable $\bm{z}(\bm{s},t)$ is based on the maximum likelihood approach considering profile data observed at spatial locations $\mathcal{S}=\{\bm{s}_{i},i=1,...,n\}$ and time points $t=1,...,T$.

At a specific location $\bm{s}_{i}$ and time $t$, $q_{i,t}$ measurements are taken at points $\bm{h}_{\bm{s}_{i},t}=\left(  h_{i,1,t},...,h_{i,q_{i,t},t}\right)  ^{\top}$ and collected in the vector
\[
\bm{y}_{\bm{s}_{i},t}=(y(\bm{s}_{i},h_{i,1,t},t),...,y(\bm{s}_{i},h_{i,q_{i,t},t},t))^{\top},
\]
here called the observed profile. 

Although \pkg{D-STEM} v2 allows for varying $q_{i,t}$, for ease of notation, it is assumed here that
all profiles include exactly $q$ measurements, although $\bm{h}_{\bm{s}_{i},t}$ may be different across profiles. 
Profiles observed at time $t$ across spatial locations $\mathcal{S}$ are then stored in the $nq\times1$
vector $\bm{y}_{t}=(\bm{y}_{s_{1},t}^{\top},...,\bm{y}_{s_{n},t}^{\top})^{\top}$. 
Applying model (\ref{eq:model_line1})-(\ref{eq:model_line3}) to the defined data above, we have the following matrix representation:
\begin{align*}
\bm{y}_{t}  &  =\tilde{\bm{X}}_{t}\bm{c}_{\bm{\beta}}+\bm{\Phi}_{\bm{z},t}\bm{z}%
_{t}+\bm{\varepsilon}_{t},\\
\bm{z}_{t}  &  =\tilde{\bm{G}}\bm{z}_{t-1}+\bm{\eta}_{t},
\end{align*}
where $\tilde{\bm{X}}_{t}=\bm{X}_{t}\bm{\Phi}_{\bm{\beta},t}$ is a $nq\times bp$
matrix, with $\bm{X}_{t}$ the matrix of covariates and $\bm{\Phi}_{\bm{\beta},t}$
the basis matrix for $\bm{\beta}$. $\bm{\Phi}_{\bm{z},t}$ is the $nq\times np$ basis
matrix for the latent $np\times1$ vector 
$\bm{z}_{t}=(\bm{z}(\bm{s}_{1},t)^{\top},...,\bm{z}(\bm{s}_{n},t)^{\top})^{\top}$. 
$\bm{\eta}_{t}=(\bm{\eta
}(\bm{s}_{1},t)^{\top},...,\bm{\eta}(\bm{s}_{n},t)^{\top})^{\top}$ is the $np\times1$ innovation vector, while $\bm{\varepsilon}_{t}\ $ is the $nq\times1$ vector of measurement errors. 
Additionally, $\tilde{\bm{G}}= \bm{I}_{n} \otimes \bm{G}$ is the $np\times np$
diagonal transition matrix.

The complete-data likelihood function $L(\bm{\psi};\bm{Y},\bm{Z})$ can
be written as
\[
L(\bm{\psi};\bm{Y},\bm{Z})=L(\bm{\psi}_{\bm{z}_{0}};\bm{z}%
_{0})\prod_{t=1}^{T}L(\bm{\psi}_{\bm{y}};\bm{y}_{t}|\bm{z}%
_{t})L(\bm{\psi}_{\bm{z}};\bm{z}_{t}|\bm{z}_{t-1}),
\]
where $\bm{Y}=\left(  \bm{y}_{1},...,\bm{y}_{T}\right)  $,
$\bm{Z}=\left(  \bm{z}_{0},\bm{z}_{1},...,\bm{z}_{T}\right)
$, $\bm{\psi}_{\bm{z}}=\left(  \bm{g}^{\top},\bm{v}^{\top},\bm{\theta}^{\top}\right) ^{\top} $, $\bm{\psi}_{\bm{y}}=\left(
\bm{c}_{\varepsilon}^{\top},\bm{c}_{\beta}^{\top}\right) ^{\top} $, and $\bm{z}_{0}$ is the Gaussian initial vector with parameter $\bm{\psi}_{\bm{z}_{0}}$. Maximum likelihood estimation is based on an extension of the EM algorithm detailed in
\cite{calculli2015maximum}. The model parameter set $\bm{\psi}$ is initialised
with starting values $\bm{\psi}^{\left\langle 0\right\rangle }$ and then
updated at each iteration $\iota$ of the EM algorithm. 

The algorithm terminates if any of the following conditions is satisfied:
\[
\max_{l}\left\vert \psi _{l}^{\left\langle \iota \right\rangle }-\psi
_{l}^{\left\langle \iota -1\right\rangle }\right\vert /\left\vert \psi
_{l}^{\left\langle \iota \right\rangle }\right\vert <\epsilon _{1}
\]
\[ \left\vert
L(\bm{\psi}^{\left\langle \iota\right\rangle };\bm{Y}%
)-L(\bm{\psi}^{\left\langle \iota-1\right\rangle };\bm{Y})\right\vert
/\left\vert L(\bm{\psi}^{\left\langle \iota\right\rangle };\bm{Y})\right\vert
<\epsilon_2,\]  
\[ \iota>\iota^{\ast}, \]
where $\psi _{l}^{\left\langle \iota \right\rangle }$ is the generic element of $\bm{\psi}^{\left\langle \iota \right\rangle }$ at the
$\iota\text{-}th$ iteration, $L(\bm{\psi}^{\left\langle \iota\right\rangle };\bm{Y})$ is the
observed-data likelihood function evaluated at $\bm{\psi}^{\left\langle
\iota\right\rangle }$, $0<\epsilon_1\ll1$ and $0<\epsilon_2\ll1$ are small positive numbers

(e.g. $10^{-4}%
$), while $\iota^{\ast}$ is a user-defined positive integer number (e.g. $100$) to limit the
iterations in the case of convergence failure of the EM algorithm.

Note that $\mathcal{S}$ is not time-varying, which means that spatial locations are fixed. This could be a limit in applications where spatial locations change for each $t$. On the other hand, missing profiles are allowed; that is, $\bm{y}_{\bm{s}_{i},t}$ may be a vector of $q$ missing values at some $t$. In the extreme case, a given spatial location $\bm{s}_{i}$ has only one profile over the entire period (if all the profiles are missing, the spatial location can be dropped from the data set).
\citet[p.~348]{shumway2017} explains how the likelihood function of a state-space model changes in the case of a missing observation vector and how the EM estimation formulas are derived. Missing data handling in \pkg{D-STEM} v2 is based on the same approach.

\subsection{Partitioning}\label{sec:partitioning}

At each iteration of the EM algorithm, the computational complexity of the E-step is
$O\left(  Tn^{3}p^{3}\right)  $, which may be unfeasible if $n$ is large.
When necessary, \pkg{D-STEM} v2 allows one to use a partitioning approach \citep{stein2013} for model
estimation. 
The spatial locations $\mathcal{S}$ are divided into $k$ partitions, 
and $\bm{z}_{t}$ is partitioned conformably, namely, $\bm{z}_{t}=\left(  \bm{z}_{t}^{(1)\top},...,\bm{z}%
_{t}^{(k)\top}\right)  ^{\top}$. 
Hence, the likelihood function becomes
\[
{\prod\limits_{t=1}^{T}}
L\left(  \bm{\psi}_{\bm{y}};\bm{y}_{t}\mid\bm{z}_{t}\right)  \cdot%
{\prod\limits_{j=1}^{k}}
L\left(  \bm{\psi}_{\bm{z}_{0}};\bm{z}_{0}^{(j)}\right)  \cdot%
{\prod\limits_{j=1}^{k}}
{\prod\limits_{t=1}^{T}}
L\left(  \bm{\psi}_{\bm{z}};\bm{z}_{t}^{(j)}\mid\bm{z}_{t-1}%
^{(j)}\right).
\]
From the EM algorithm point of view, this implies
that the E-step is independently applied to each partition, possibly in parallel. When all partitions are equal in size, 
the computational complexity reduces to $\mathcal{O}\left( Tkr^{3}p^{3}\right)$, with $r$ as the partition size.

Geographical partitioning, constructed aggregating proximal locations, is a natural choice for environmental applications.
Given the number of partitions $k$, the k-means algorithm applied to spatial coordinates
provides a geographical partitioning of $\mathcal{S}$. 
However, the number of points in each partition is not controlled, and a heterogeneous partitioning may arise.
If some subsets are very large and others are small, the reduction in computational complexity given above is far from being achieved.
This can easily happen, for example, when $\mathcal{S}$ is a global network constrained by continent shapes.

For this reason, \pkg{D-STEM} v2 provides a heuristically modified k-means algorithm that encourages 
partitions with similar numbers of elements. 
The algorithm optimises the following objective function:

\begin{equation}
\sum_{j=1}^{k}\sum_{\bm{s}\in\mathcal{S}_{j}}d\left(  \bm{s},\bm{c}_{j}\right)  +\lambda\sum_{j=1}^{k}\left(
r_{j}-\frac{n}{k}\right)  ^{2}, \label{eq:k-means}%
\end{equation}

where $\lambda\ge0$, $\mathcal{S}_{j} \subset \mathcal{S}$ is the set of coordinates in the $j\text{-}th$ 
partition, $d$ is the geodesic distance on the sphere $\mathbb{S}^{2}$ and $\bm{c}_{j}$ and $r_{j}$ 
are the centroid and the number of elements in the $j\text{-}th$ partition, respectively. \\
The second term in (\ref{eq:k-means}) accounts for the variability of the partition sizes and acts as a penalisation for heterogeneous partitionings.
Clearly, when $\lambda=0$, the above-mentioned objective function gives the classic k-means algorithm. 
For high values of $\lambda$, solutions with similarly sized partitions are favoured. \\
Unfortunately, an optimality theory for this algorithm has not yet been developed, and the choice of $\lambda$ is left to the user. Nonetheless, it may be a useful tool to define a partitioning that is appropriate for the application at hand with regard to computing time and geographical properties.

\subsection{Variance-covariance matrix estimation}\label{sec:varcov}

The EM algorithm provides a point estimate of the parameter vector $\bm{\psi}$
but no uncertainty information. Building on \citet[p.~408]{shumway2017}, \pkg{D-STEM} v2
estimates the variance--covariance matrix 
$\Sigma_{\bm{\psi},T}=\mathbb{V}\left(\bm{\psi}\mid\bm{Y}\right)$, 
by means of the observed Fisher information matrix, $\mathbf{I}_{T}$, namely 
$$\hat{\Sigma}_{\bm{\psi},T}=(\mathbf{I}_{T})^{-1}.$$
To understand its computational cost, note that the information matrix given above may be written as a sum:
$\mathbf{I}_{T}=\sum_{t=1}^T\mathbf{i}_t$.\\
For large data sets, each matrix $\mathbf{i}_t$ may be expensive to compute, and the total computational cost is linear in $T$, provided missing data are evenly distributed in time.
This results in a time-consuming task with a computational burden even higher than that for model estimation.
For this reason, \pkg{D-STEM} v2 makes it possible to approximate $\hat{\Sigma}_{\bm{\psi},T}$ using a truncated information matrix, namely:
\begin{equation}
\tilde{\Sigma}_{\bm{\psi},t^*}=(\frac{T}{t^*}\mathbf{I}_{t^*})^{-1},
\label{eq:Fisher_approximated}
\end{equation}
which reduces the computational burden by a factor of $1-t^*/T$.\\
Since
$\tilde{\Sigma}_{\bm{\psi},t^*} \rightarrow \hat{\Sigma}_{\bm{\psi},T}$ for $t^* \rightarrow T$,
the truncation time $t^*$ is chosen to control the approximation error in $\hat{\Sigma}_{\bm{\psi}}$. In particular, $t^*$ is the first integer such that
\begin{equation}
\frac{\left\Vert \tilde{\Sigma}_{\bm{\psi},t}-\tilde{\Sigma}_{\bm{\psi},t-1}%
\right\Vert_{F} }{\left\Vert \tilde{\Sigma}_{\bm{\psi},t}\right\Vert_{F} }\leq
\delta,\label{eq:varcov_approximated}
\end{equation}
where $\left\Vert { \cdot }\right\Vert_{F}$ is the Frobenius norm, and $\delta$ may be defined by the user.\\
Generally speaking, the behaviour of $\hat{\Sigma}_{\bm{\psi},T}$ for large $T$ and, hence, the behaviour of $\tilde{\Sigma}_{\bm{\psi},t}$ relays on stationarity and ergodicity of the underlying stochastic process; see, for example, 
\citet[Property P6.4]{shumway2017} and references therein.\\
To have operative guidance for the user, let us assume first that no missing values are present, the information matrix is well-conditioned and the covariates have no isolated outliers or extreme trends. 
In this case, away from the borders $t\cong1$ and $t\cong T$, 
the observed conditional information $\mathbf{i}_t$ has a relatively smooth stochastic behaviour, and the approximation in (\ref{eq:Fisher_approximated}) is expected to be satisfactory at the level defined by $\delta$.
Conversely, if some data are missing at time $t$, the information $\mathbf{i}_t$ is reduced accordingly. If the missing pattern is random over time, this is not an issue. 
But, in the unfavourable case with a high percentage of missing data mostly concentrated at the end the time series, $t\cong T$, the above approximation may over-estimate the information and under-estimate the variances of the parameter estimates.

\subsection{Dynamic kriging}\label{sec:kriging}

In this paper, dynamic kriging refers to evaluating the following
quantities:

\begin{align}
\hat{f} \left(  \bm{s},h,t\right) &= \mathbb{E}_{\hat{\bm{\psi}}}\left(  f\left(  \bm{s},h,t\right)  \mid
\bm{Y}\right), \label{eq:krig_exp}\\
\VAR\left( \hat{f} \left(  \bm{s},h,t\right) \right) &= \mathbb{V}_{\hat{\bm{\psi}}}\left(  f\left(  \bm{s},h,t\right)  \mid\bm{Y}\right),  \label{eq:krig_var}%
\end{align}
\\
for any $\bm{s}\in\mathbb{S}^{2}$, $h\in\mathcal{H}$ and $t=1,...,T$. A common approach is to map the kriging estimates on a regular pixelation $\mathcal{S}^{\ast}=\left\{  \bm{s}_{1}^{\ast},...,\bm{s}_{m}^{\ast}\right\}  $. This may be a time-consuming task when $m$ and/or $n$ and/or $T$ are large. To tackle this problem, \pkg{D-STEM} v2 allows one to exploit a nearest-neighbour approach, where the conditioning term in Equations (\ref{eq:krig_exp}) and
(\ref{eq:krig_var}) is not $\bm{Y}$, but the data at the spatial locations $\mathcal{S}_{\sim j}$, where
$\mathcal{S}_{\sim j}\subset\mathcal{S}$ is the set of the $\tilde{n}\ll n$ nearest
spatial locations to $\bm{s}_{j}^{\ast}$. The use of the nearest-neighbour approach
is justified by the so-called screening effect. Even when the spatial correlation function exhibits 
long-range dependence, it can subsequently be assumed that $y$ at spatial location $\bm{s}$ is nearly independent of spatially distant
observations when conditioned on nearby observations \citep[see][for more details]{Stein2002,furrer2006}.

For computational efficiency, \pkg{D-STEM} v2 performs kriging for blocks of pixels. 
To do this, $\mathcal{S}^{\ast}$ is partitioned in $u$ blocks $\mathcal{S}^{\ast}=\left\{\mathcal{S}_{1}^{\ast},...,\mathcal{S}_{u}^{\ast}\right\}$, and kriging is
done on each block $\mathcal{S}_{l}^{\ast}$, $l=1,...,u$, with $u\ll m$ controlled by the user.
For each target block $\mathcal{S}_{l}^{\ast}$, the conditioning term in Equations
(\ref{eq:krig_exp}) and (\ref{eq:krig_var}) is given by the data
observed at $\mathcal{\tilde{S}}_{l}=\bigcup\nolimits_{j\in \mathcal{J}_{l}}\mathcal{S}_{\sim j},%
\mathcal{J}_{l}=\left\{ j:s_{j}^{\ast }\in \mathcal{S}_{l}^{\ast}\right\}$. 
Note that, if $\mathcal{S}_{l}^{\ast}$ is dense and $\mathcal{S}$ is sparse (namely $n\ll m$), then $\mathcal{\tilde{S}}_{l}$ is not 
much larger than $\mathcal{S}_{\sim j}$ since most of the spatial locations in $\mathcal{S}_{l}^{\ast}$ 
tend to have the same neighbours $\mathcal{S}_{\sim j}$.

\subsection{Validation}\label{sec:val}

\pkg{D-STEM} v2 allows one to implement an out-of-sample validation by partitioning the original spatial locations
$\mathcal{S}$ into subsets $\mathcal{S}_{est}$ and $\mathcal{S}_{val}$. Data at $\mathcal{S}%
_{est}$ are used for model estimation while data at $\mathcal{S}%
_{val}$ are used for validation. Once the model is estimated, the kriging
formula in Equation (\ref{eq:krig_exp}) is used to predict at $\mathcal{S}_{val}$ for all times $t$ and heights $\bm{h}$. The following validation mean squared errors are
then computed 
\begin{align*}
MSE_{t}  &  =\frac{1}{P_{1}}\sum_{\bm{s}\in\mathcal{S}_{val}}%
\sum_{h\in\bm{h}_{\bm{s},t}}\left(  y\left(  \bm{s},h,t\right)  -\hat{y}\left(  \bm{s},h,t\right)  \right)  ^{2},\\
MSE_{\bm{s}}  &  =\frac{1}{P_{2}}\sum_{t=1}^{T}\sum_{h\in
\bm{h}_{\bm{s},t}}\left(  y\left(  \bm{s},h,t\right)
-\hat{y}\left(  \bm{s},h,t\right)  \right)  ^{2},\\
MSE_{h}  &  =\frac{1}{P_{3}}\sum_{t=1}^{T}\sum_{\bm{s}\in\mathcal{S}_{val}}\left(  y\left(  \bm{s},h,t\right)
-\hat{y}\left(  \bm{s},h,t\right)  \right)  ^{2},%
\end{align*}
where $\hat{y}\left(  \bm{s},h,t\right)  $ is obtained from Equation (\ref{eq:krig_exp}), while $P_{1}$, $P_{2}$ and $P_{3}$ are the number of
terms in each sum. 

When $\bm{h}_{\bm{s},t}$ varies across the profiles, \pkg{D-STEM} v2 provides a binned MSE by splitting the continuous domain $\mathcal{H}$ into $B$ equally spaced intervals. Let $H^*_r$ be the set of observation points in the $r\text{-}th$ interval, let $n_r$ be the corresponding observation number and let $\bar{h}_r = \frac{1}{n_r} \sum_{h\in H^*_r} h$ be the mean of points in $b\text{-}th$ interval. Then, the $MSE_{\bar{h}_r}$ is computed by
\begin{align*}
MSE_{\bar{h}_r}  &  =\frac{1}{P_{4}} \sum_{h\in H^*_r} \sum_{t=1}^{T}\sum_{\bm{s}\in\mathcal{S}_{val}}\left(  y\left(  \bm{s},h,t\right)
-\hat{y}\left(  \bm{s},h,t\right)  \right)  ^{2},%
\end{align*}
where $P_{4}$ is the total number of observations in the $b\text{-}th$ interval. 

D-STEM v2 also provides the validation $R^2$ with respect to time
\begin{align*}
R^{2}_{t}  &  =1 - \frac{MSE_{t}}{\VAR\left( \{y\left(  \bm{s},h,t\right), \bm{s}\in\mathcal{S}_{val}, h\in\bm{h}_{\bm{s},t} \} \right)}.
\end{align*}
and the analogous validation $R^2$ with respect to location $\bm{s}$ and $h_r$.

\section[Software]{Software}\label{sec:software}

This section starts by briefly describing the modelling capabilities of \pkg{D-STEM} v2 inherited by the previous version for dealing with spatio-temporal data sets. Then, it focuses on the \pkg{D-STEM} v2 classes and methods, which implement estimation, validation and dynamic mapping of the model presented in Section \ref{sec:methodology}. Although some of the classes are already available in \pkg{D-STEM} v1, they are listed here for completeness.

\subsection{Software description}

\pkg{D-STEM} v1 implemented a substantial number of models. The dynamic coregionalisation model (DCM, \citeauthor{finazzi2014d} \citeyear{finazzi2014d}) and the hidden dynamic geostatistical model (HDGM, \citeauthor{calculli2015maximum} \citeyear{calculli2015maximum})
are suitable for modelling and mapping multivariate space-time data collected from
unbalanced monitoring networks. Model-based clustering (MBC, \citeauthor{finazzi2015} \citeyear{finazzi2015})  has been introduced
for clustering time series, and it is suitable for large data sets with spatially registered time series.
Moreover, the emulator model \citep{finazzi2019} is based on a Gaussian emulator, and it is exploited for
modelling the multivariate output of a complex physical model.

In addition, \pkg{D-STEM} v2 (available at \url{github.com/graspa-group/d-stem}) provides the functional version of HDGM, denoted by f-HDGM, which handles modelling and mapping of functional space-time data, following the methodology of Section \ref{sec:methodology}. For implementing f-HDGM,
\pkg{D-STEM} v2 relies on the \proglang{MATLAB} version of the \pkg{fda} package \citep{ramsay2018}, which is automatically downloaded and installed by \pkg{D-STEM} v2.

\subsection{Data format}\label{sec:data.format}

Two data formats are available to define observations for the f-HDGM. One is the internal format used by the \pkg{D-STEM} v2 classes, and the other one is the user format based on the more user-friendly \code{table} data type implemented in recent versions of \proglang{MATLAB}.
The latter permits storing measurement profiles, covariate profiles, coordinates, timestamps and units of measure in a single object. The internal format is not discussed here.\\
Considering a table in the user format, each row includes the profiles collected at a given spatial location and time point. 
The column labels are defined as follows:
columns \code{Y} and \code{Y\_name} are used for the dependent variable $y$ and its name as a string field, respectively;
the column with prefix \code{X\_h\_} is used for the values of the domain $h$; eventually, columns with prefix \code{X\_beta\_} are used for covariates $\bm{x}$. 
These tables have only one column for $y$ and only one column for $h$. Instead, we can have any number $b \geqslant 0$ of covariate columns. Additionally, the table has columns \code{X\_coordinate} and \code{Y\_coordinate} for  spatial location $\bm{s}$ and column \code{Time} for the timestamp. Units of measure are stored in the \code{Properties.VariableUnits} property of the table columns and used in outputs and plots. Units for \code{X\_coordinate} and \code{Y\_coordinate} can be \code{deg} for degrees, \code{m} for meters and \code{km} for kilometres. Geodetic distance is used when the unit is \code{deg}; otherwise, the Euclidean distance is used.\\
At the table row corresponding to location $\bm{s}_i$ and time $t$, the elements related to $y$ and $\bm{x}$ are vectors with $q_{i,t}$ elements. 
Vectors related to $y$ may include missing data (\code{NaN}). If $y$ is entirely missing for a
given $\left(  \bm{s},t\right)  $, the row must be removed from the table.
Since spatial locations $\mathcal{S}$ are fixed in time, and as their number $n$ is
determined by the number of unique coordinates in the table, 
profiles observed at different time points but the same spatial location $\bm{s}$ must have
the same coordinates.\\

\subsection{Software structure}

In \pkg{D-STEM}, a hierarchical structure of object classes and methods is used to handle data definition, model definition and estimation, validation, dynamic kriging and the related plotting capabilities. 
The structure is schematically given below. 
Further details on the use of each class are given within the two case studies in this paper, while class constructors, methods and property details can be obtained in \proglang{MATLAB} using the command

\code{doc <class_name>}.

\subsubsection{Data handling}
The \code{stem_data} class allows the user to define the data used in f-HDGM models, mainly through the following objects and methods.
\begin{itemize}
\item Objects of \code{stem_data}
\begin{itemize}
\item \code{stem_modeltype}: model type (DCM, HDGM, MBC, Emulator or f-HDGM); note that model type is needed here because the data structure varies among the different models;
\item \code{stem_fda}: basis functions specification;
\item \code{stem_validation} (optional): definition of the learning and testing datasets for model validation.
\end{itemize}
\item Methods and Properties of \code{stem_data}
\begin{itemize}
\item \code{kmeans_partitioning}: data partitioning for parallel EM computations of Section \ref{sec:partitioning}; 
this method is applied to a \code{stem_data} object, and its output is used by the \code{EM_estimation} method in the \code{stem_model} class below;
\item \code{shape} (optional): structure with geographical borders used for mapping.
\end{itemize}
\item Internal Objects of \code{stem_data}
\begin{itemize}
\item \code{stem_varset}: observed data and covariates;
\item \code{stem_gridlist}: list of \code{stem_grid} objects
\begin{itemize}
\item \code{stem_grid}: spatial locations coordinates;
\end{itemize}
\item \code{stem_datestamp}: temporal information.
\end{itemize}
\end{itemize}

Interestingly, \code{stem_misc.data_formatter} is a helper method, which is useful for building \code{stem_\-varset} objects starting from data tables. Its class,  \code{stem_misc}, is a miscellanea static class implementing other methods for various intermediate tasks not discussed here for brevity.
\subsubsection{Model building}

The \code{stem_model} class is used to define, estimate, validate and output a f-HDGM, mainly through the following objects and methods.\\
\begin{itemize}
\item Objects of \code{stem_model}
\begin{itemize}
\item \code{stem_data}: defined above;
\item \code{stem_par}: model parameters;
\item \code{stem_EM_result}: container of the estimation output, after \code{EM_estimate};
\item \code{stem_validation_result} (optional): container of validation output, available only if \code{stem_data} contains the \code{stem_validation} object;
\item \code{stem_EM_options} (optional): model estimation options; it is an input of the  \code{EM_estimate} method below.
\end{itemize}
\item Methods of \code{stem_model}
\begin{itemize}
\item \code{EM_estimate}: computation of parameter estimates; 
\item \code{set_varcov}: computation of the estimated variance-covariance matrix;
\item \code{plot_profile}: plot of functional data;
\item \code{print}: print estimated model summary;
\item \code{beta_Chi2_test}: testing significance of covariates;
\item \code{plot_par}: plot functional parameter;
\item \code{plot_validation}: plot MSE validation.
\end{itemize}
\end{itemize}
\subsubsection{Kriging}
The kriging handling is implemented with two classes. The first is the \code{stem_krig} class, which implements the kriging spatial interpolation.
\begin{itemize}
\item Objects of \code{stem_krig}
\begin{itemize}
\item \code{stem_krig_data}: mesh data for kriging;
\item \code{stem_krig_options}: kriging options;
\end{itemize}
\end{itemize}
\begin{itemize}
\item Methods of \code{stem_krig}
\begin{itemize}
\item \code{kriging}: computation of kriging, the output is a \code{stem_krig_result} object.
\end{itemize}
\end{itemize}
The second is the \code{stem_krig_result} class, which stores the kriging output and implements the methods for plotting the kriging output.
\begin{itemize}
\item Methods of \code{stem_krig_result}
\begin{itemize}
\item \code{surface_plot}: mapping of kriging estimate and their standard deviation for fixed $h$;
\item \code{profile_plot}: method for plotting the kriging function and the variance-covariance matrix for a fixed space and time. 
\end{itemize}
\end{itemize}
Although at first reading the user could prefer a single object for both input and output of the kriging, these objects may be quite large, making the current approach more flexible.

\section[Case study on ozone data]{Case study on ozone data}\label{sec:casestudy_ozone}

This section illustrates how to make inferences on an f-HDGM for ground-level high-frequency air
quality data collected by a monitoring network. In particular, hourly ozone ($O_{3}$, in $\mu g/m^3$) measured
in Beijing, China, is considered.

\subsection{Air quality data}

Ground-level $O_{3}$ is an increasing public concern due to its
essential role in air pollution and climate change. In China, $O_{3}$ has
become one of the most severe air pollutants in recent years
\citep{wang2017ozone}.

\begin{figure}[ptb]
\centering
\includegraphics[width=0.63\textwidth]{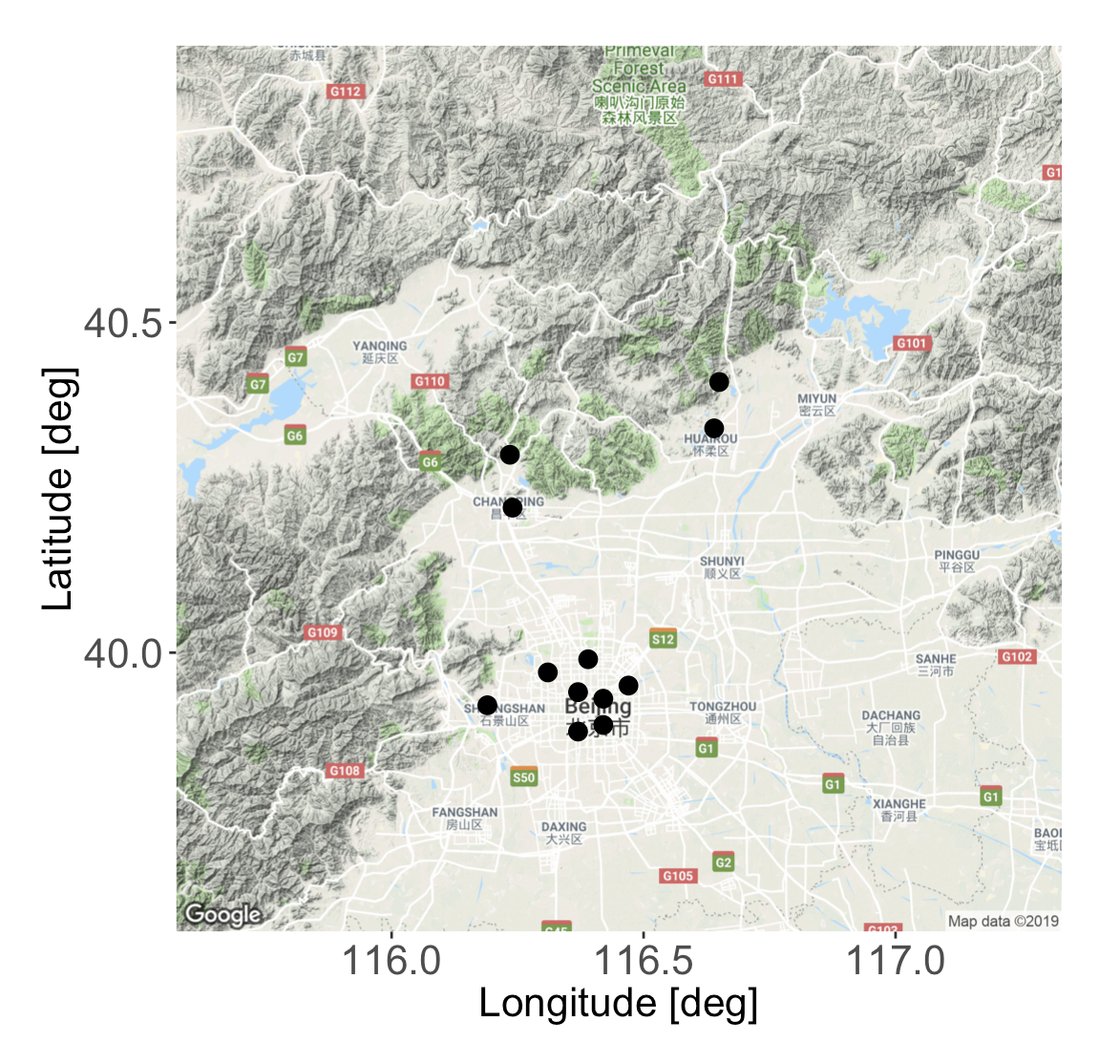} \caption{Spatial
locations of the twelve stations in Beijing \citep{david2013ggmap}.}%
\label{fig:mapsite}%
\end{figure}

In this case study, the aim is to model hourly $O_{3}$ concentrations from 2015 to 2017 with respect to temperature and ultraviolet radiation (UVB) across Beijing.
Concentration and temperature data are available at twelve monitoring stations (Figure \ref{fig:mapsite}). Hourly UVB data are obtained from the ERA-Interim product of the European Centre for Medium-Range Weather Forecasts (ECMWF) at a grid size of $0.25^{\circ} \times 0.25^{\circ}$ over the city.

To describe the diurnal cycle of $O_{3}$, which peaks in the afternoon and reaches a minimum at night-time, the 24 hours of the day are used as domain $\mathcal{H}$ of the basis functions, while the time index $t$ is on the daily scale. Moreover, due to the circularity of
time, Fourier basis functions are adopted, which implies that $\beta_{j}\left(
h\right)  $, $\sigma_{\varepsilon}^{2}\left(  h\right)  $ are periodic functions.

The measurement equation for $O_{3}$ is
\begin{equation}
y(\bm{s},h,t)=\beta_{0}(h)+x_{temp}(\bm{s},h,t)\beta
_{temp}(h)+x_{uvb}\left(  t\right)  \beta_{uvb}(h)+\bm{\phi}(h)^{\top}\bm{z}(\bm{s},t)+\varepsilon(\bm{s},h,t), \label{eq:O3meaurement}%
\end{equation}
where $\bm{s}$ is the generic spatial location, $h\in\left[  0,24\right)
$ is the time within the day expressed in hours and $t=1,...,1096$ is the day
index over the period 2015--2017. Based on a preliminary analysis, the number of basis functions for $\beta_{j}\left(  h\right)  $, $\sigma_{\varepsilon}^{2}\left(  h\right)  $
and $\bm{\phi}(h)^{\top}\bm{z}(\bm{s},t)$ is chosen to be $5$, $5$
and $7$, respectively.

\subsection{Software implementation}

This paragraph details the implementation of the \pkg{D-STEM} v2 in three
aspects: model estimation, validation and kriging. Relevant
scripts are \code{demo_section4_model_estimate.m},
\code{demo_section4_validation.m} and \code{demo_section4_kriging.m},
respectively, which are available in the supplementary material. All the scripts can be executed by choosing the option number from 1 to 3 in the \code{demo_menu_user.m} script.

\subsubsection{Model estimation}

This paragraph describes the \code{demo_section4_model_estimate.m} script
devoted to the estimation of the model parameters and of their
variance--covariance matrix.

The data set needed to perform this case study is stored as a \proglang{MATLAB} table in the user format of Section \ref{sec:data.format} and named \code{Beijing_O3}. It can be loaded from the corresponding file as follows:

\begin{CodeChunk}
\begin{CodeInput}
load ../Data/Beijing_O3.mat;
\end{CodeInput}
\end{CodeChunk}

In the \code{Beijing_O3} table, each row refers to a fixed space-time point and gives a 24-element hourly ozone profile with the corresponding conformable covariates, which are: a constant, temperature and UVB.

The following lines of code specify the model type and the basis functions, which are stored in an object of class \code{stem_fda}:

\begin{CodeChunk}
\begin{CodeInput}
o_modeltype = stem_modeltype('f-HDGM');
input_fda.spline_type = 'Fourier';
input_fda.spline_range = [0 24];
input_fda.spline_nbasis_z = 7;
input_fda.spline_nbasis_beta = 5;
input_fda.spline_nbasis_sigma = 5;
o_fda = stem_fda(input_fda);
\end{CodeInput}
\end{CodeChunk}

When using a Fourier basis, \code{spline_nbasis_z} must be set to a positive odd
number. 
Meanwhile, \code{spline_nbasis_beta} and/or \code{spline_nbasis_sigma} must be left empty, if $\bm{\beta}(h)\equiv\bm{\beta}$ and/or $\sigma
_{\varepsilon}^{2}(h)\equiv\sigma_{\varepsilon}^{2}$ are constant functions.

The next step is to define an object of class \code{stem_data}, which specifies the model type and contains the basis function object and the data from the \code{Beijing_O3} table, transformed in the internal data format. This is done using the intermediate \code{input_data} structure:

\begin{CodeChunk}
\begin{CodeInput}
input_data.stem_modeltype = o_modeltype;
\end{CodeInput}
\vspace{-.95cm} 
\begin{CodeInput}
input_data.data_table = Beijing_O3;
\end{CodeInput}
\vspace{-.95cm} 
\begin{CodeInput}
input_data.stem_fda = o_fda;
o_data = stem_data(input_data);
\end{CodeInput}
\end{CodeChunk}

Then, an object of class \code{stem_model} is created by using both information on data, stored in the \code{o_data} object, and on parametrisation, contained in the \code{stem_par} object named \code{o_par}:

\begin{CodeChunk}
\begin{CodeInput}
o_par = stem_par(o_data,'exponential');
o_model = stem_model(o_data, o_par);
\end{CodeInput}
\end{CodeChunk}

\begin{figure}[htp]
\centering
\includegraphics[width=0.75\textwidth]{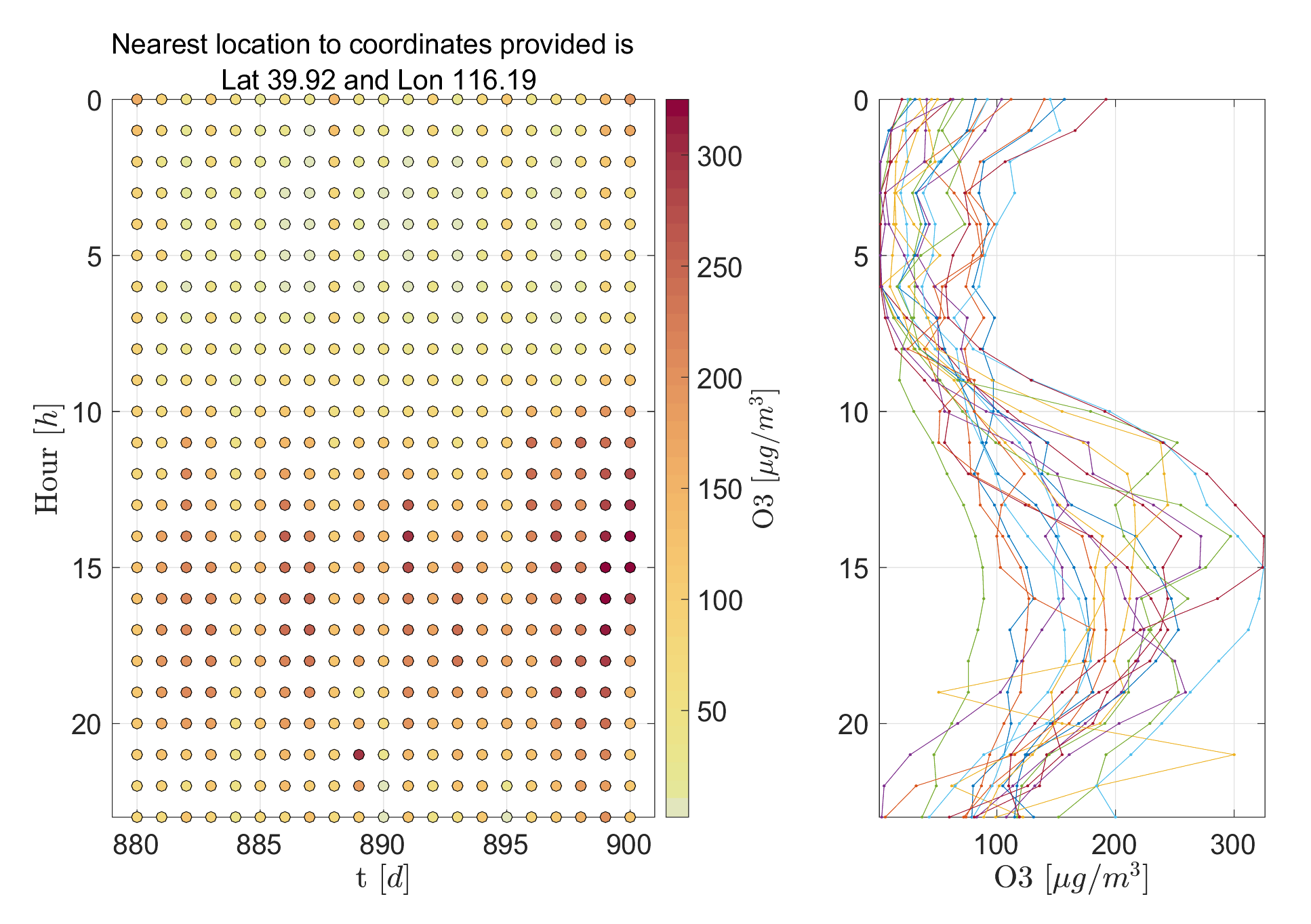}
\caption{\label{fig:data_profile_o3} $O_3$ concentrations at location 39.92 latitude and 116.19 longitude for 21 days beginning on 29 May 2017. Left: each dot is a concentration measurement. The colour of the dot depicts the concentration. Right: each graph is a daily concentration profile.}
\end{figure}

To facilitate visualisation, the method \code{plot_profile} of class \code{stem_model} shows the $O_3$ profile data at location (\code{lat0}, \code{lon0}), in the days between \code{t_start} and \code{t_end} (Figure \ref{fig:data_profile_o3}):

\begin{CodeChunk}
\begin{CodeInput}
lat0 = 40; lon0 = 116;
t_start = 880; t_end = 900;
o_model.plot_profile(lat0, lon0, t_start, t_end);
\end{CodeInput}
\end{CodeChunk}

Before running the EM algorithm, the model parameters need to be initialised.
This is done using the method \code{get_beta0} of class \code{stem_model}, which
provides the starting values for $\bm{\beta}$, and the method
\code{get_coe_log_sigma_eps0} for the case of a functional $\sigma_{\varepsilon}^{2}(h)$.
Next, the method \code{set_initial_values} of the \code{o_model} object is called to complete the initialisation of model parameters:

\begin{CodeChunk}
\begin{CodeInput}
n_basis = o_fda.get_basis_number;
o_par.beta = o_model.get_beta0();
o_par.sigma_eps = o_model.get_coe_log_sigma_eps0();
o_par.theta_z = ones(1, n_basis.z)*0.18;
o_par.G = eye(n_basis.z)*0.5;
o_par.v_z = eye(n_basis.z)*10;
o_model.set_initial_values(o_par);	
\end{CodeInput}
\end{CodeChunk}

Note that the \code{theta_z} parameter must be provided in the same unit of measure as the spatial coordinates.

Before model estimation, EM exiting conditions $\epsilon_1$ (\code{exit_toll_par}), $\epsilon_2$ (\code{exit_toll_loglike}) and $\iota^{\ast}$ (\code{max_iterations}) introduced in Section \ref{sec:model_estimation} can be optionally defined as follows:
\begin{CodeChunk}
\begin{CodeInput}
o_EM_options = stem_EM_options();
o_EM_options.exit_toll_par = 0.0001;
o_EM_options.exit_toll_loglike = 0.0001;
o_EM_options.max_iterations = 200;
\end{CodeInput}
\end{CodeChunk}
Model estimation is started by calling the method \code{EM_estimate} of the
\code{o_model} object, with the optional \code{o_EM_options} object passed as an input argument. 
After model estimation, the variance--covariance matrix of the estimated parameters is evaluated by calling the method \code{set_varcov}, with the optional approximation level $\delta$ of Equation (\ref{eq:varcov_approximated}) passed as an input parameter.
Finally, \code{set_logL} computes the observed data log-likelihood.

\begin{CodeChunk}
\begin{CodeInput}
o_model.EM_estimate(o_EM_options);
delta = 0.001;
o_model.set_varcov(delta);
o_model.set_logL();	
\end{CodeInput}
\end{CodeChunk}

All the relevant estimation results are found in the internal
\code{stem_EM_result} object, which can be accessed as a property of the
\code{o_model} object as follows:

\begin{CodeChunk}
\begin{CodeInput}
o_model.plot_par;
o_model.beta_Chi2_test;
o_model.print;
\end{CodeInput}
\end{CodeChunk}

\begin{figure}
\centering
\includegraphics[width=0.85\textwidth]{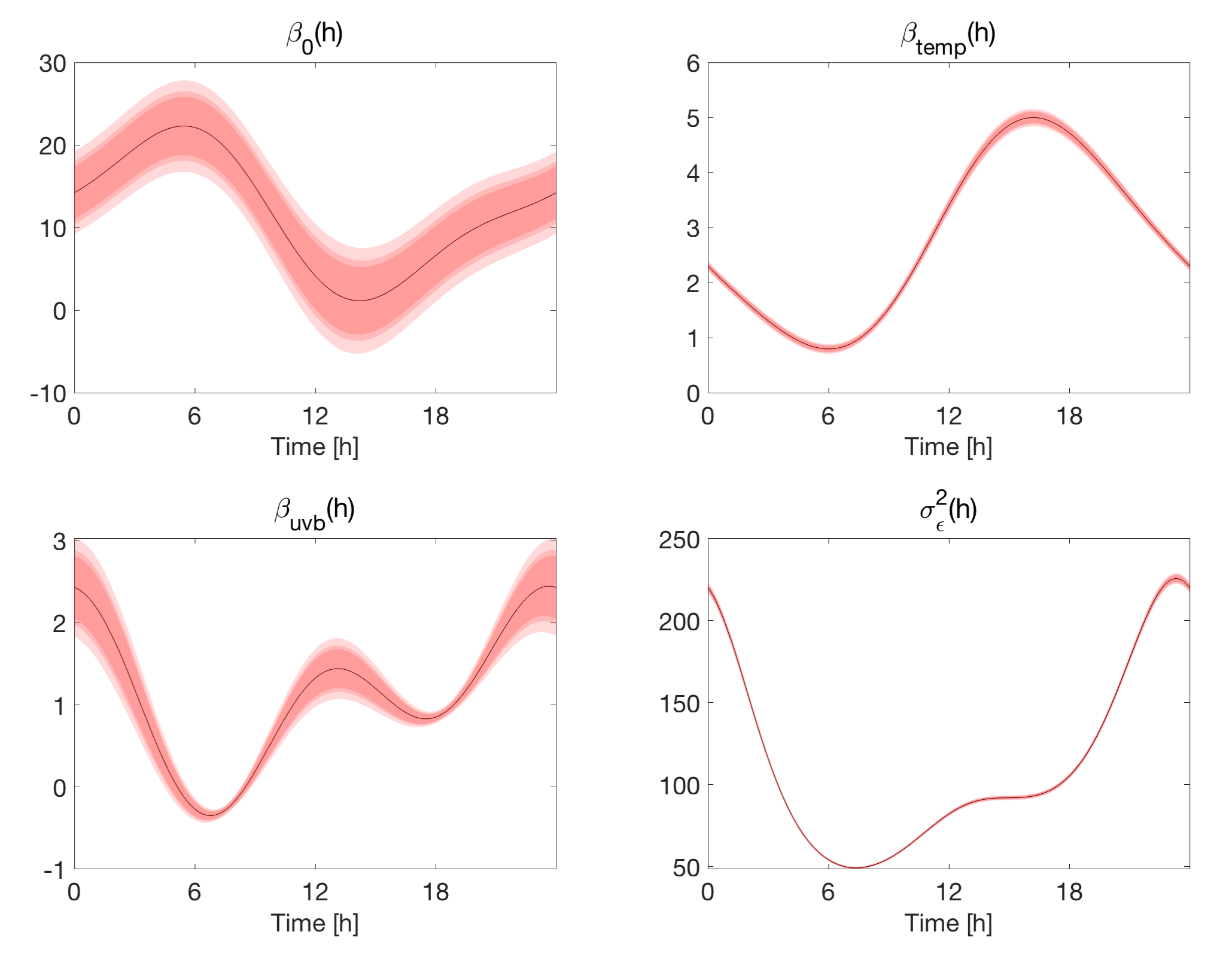} \caption{Estimated
$\beta_{0}(h),~\beta_{temp}(h),~\beta_{uvb}(h)$ and
$\sigma_{\epsilon}^{2}(h)$, with $90\%,~95\%,~99\%$ confidence bands, respectively, shown through the different shades.}%
\label{fig:par_plot_o3}%
\end{figure}

Figure \ref{fig:par_plot_o3} is produced by calling the \code{plot_par} method and 
shows the estimated $\bm{\beta}_{0}(h)$, $\bm{\beta}_{temp}(h)$, $\bm{\beta}_{uvb}(h)$, and $\sigma_{\varepsilon}^{2}(h)$. 
Thanks to the use of a Fourier basis, the functions are periodic with a period of one day. In the plot
of $\sigma_{\varepsilon}^{2}(h)$, the unexplained portion of $O_{3}$ variance,
$\sigma_{\varepsilon}^{2}(h)$, is small during daylight hours, which is
consistent with the results of \cite{dohan1987photochemical}.

When the confidence bands of \code{parplot} contain zero, it
may be useful to test the significance of the covariates. By calling the method \code{beta_Chi2_test},
the results of $\chi^{2}$ tests are obtained, and they are reported in Table \ref{tab:Chi2}. Although $\beta_{uvb}$ is
close to $0$ in the morning, all fixed effects are highly significant overall. The model output is shown in the
\proglang{MATLAB} command window by calling the \code{print} method.

\begin{table}
\centering%
\begin{tabular}
[c]{ccc}%
\toprule & $\chi^{2}$~statistic & $p$~value\\\hline
Constant & \multicolumn{1}{r}{136.33} & \multicolumn{1}{r}{0}\\\hline
Temperature & \multicolumn{1}{r}{14266.07} & \multicolumn{1}{r}{0}\\\hline
UVB & \multicolumn{1}{r}{2094.34} & \multicolumn{1}{r}{0}\\
\bottomrule &  &
\end{tabular}
\caption{$\chi^{2}$~tests for significance of covariates.}%
\label{tab:Chi2}%
\end{table}

\subsubsection{Validation}\label{sec:O3_validation}

This paragraph describes the script \code{demo_section4_validation.m}, which implements validation. Compared
to the code in \code{demo_section4_model_estimate.m}, it only differs in
providing an object of class \code{stem_validation}. 

To create the object called \code{o_validation}, the name of the validation
variable is needed as well as the indices of the validation stations.  Moreover, if the size of the nearest neighbour
set for each kriging site (\code{nn_size}) is not provided as the third input
argument in the \code{stem_validation} class constructor, \pkg{D-STEM} v2 uses all the
remaining stations. For example, a validation data set with three stations is constructed as follows:

\begin{CodeChunk}
\begin{CodeInput}
S_val = [1,7,10];
input_data.stem_validation = stem_validation('O3', S_val);
\end{CodeInput}
\end{CodeChunk}

The validation statistics, computed by \code{EM_estimate}, are saved in the internal object
\code{stem_\-validation_result}, which can be accessed as a property of the \code{o_model} object. The \code{stem_validation_result} object contains the estimated $O_{3}$ residuals for the above-mentioned validation stations as well as the validation mean square errors and $R^2$, as defined in Section \ref{sec:val}.

\subsubsection{Kriging}\label{sec:O3_kriging}

This paragraph describes the \code{demo_section4_kriging.m} script, which applies the
approach of Section \ref{sec:kriging} to 
the estimated model to map the $O_{3}$ concentrations over Beijing city.

The first step is to create an object of class
\code{stem_grid}, which collects the information about the regular grid of
pixels $\mathcal{B}$ to be used for mapping. 
Then, an object of class \code{stem_krig_data} is created, where the \code{o_krig_grid} object is passed as an input argument:

\begin{CodeChunk}
\begin{CodeInput}
load ../Output/ozone_model;
\end{CodeInput}
\vspace{-9mm}
\begin{CodeInput}
step_deg = 0.05;
lat = 39.4:step_deg:41.1;
lon = 115.4:step_deg:117.5;
[lon_mat,lat_mat] = meshgrid(lon,lat);
\end{CodeInput}
\vspace{-9mm}
\begin{CodeInput}
krig_coordinates = [lat_mat(:) lon_mat(:)];
o_krig_grid = stem_grid(krig_coordinates,'deg','regular','pixel',...
              size(lat_mat),'square',0.05,0.05);
o_krig_data = stem_krig_data(o_krig_grid);
\end{CodeInput}
\end{CodeChunk}

Two comments on the above lines follow. First, since the grid in the \code{o_krig_grid} object is regular, the dimensions of the grid (\code{size(lat_mat)}, $35 \times 43$), must be provided as well as the shape of the pixels and the spatial resolution of the grid, which is $0.05^{\circ} \times 0.05 ^{\circ}$. Second, the above step using the \code{stem_krig_data} constructor may appear redundant at first glance. Indeed, it is needed for compatibility with other model types for which, in addition to the \code{stem_grid} object, other information is also necessary for the \code{stem_krig_data} constructor.

Next, the \code{stem_krig_options} class provides some options for kriging. 
By default, the output is back-transformed in the original unit of measure if the observations have been log-transformed and/or standardised. The \code{back_\-transform} property enables handling this.
Moreover, the \code{no_varcov} property must be set to 1 to avoid the time-consuming computation of the kriging variance. 
Eventually, the \code{block_size} property is used to define the number of spatial locations in $\mathcal{S}%
_{l}^{\ast}$. 

\begin{CodeChunk}
\begin{CodeInput}
o_krig_options = stem_krig_options();
o_krig_options.back_transform = 0;
o_krig_options.no_varcov = 0;
o_krig_options.block_size = 30; 
\end{CodeInput}
\end{CodeChunk}

After storing the map of Beijing boundaries into the \code{o_model} object, the latter is used with \code{o_krig_data} to create an object of class \code{stem_krig}. 
This and \code{o_krig_options} together contain all information for kriging, which is obtained by the corresponding \code{kriging} method:
\begin{CodeChunk}
\begin{CodeInput}
o_model.stem_data.shape = shaperead('../Maps/Beijing_adm1.shp');
o_krig = stem_krig(o_model, o_krig_data);
o_krig_result = o_krig.kriging(o_krig_options);
\end{CodeInput}
\end{CodeChunk}

Note that this task may be time consuming for large grids. The kriging output saved in the \code{o_krig_result} object gives the latent process estimate $z_t$ and its variance. 
The \code{surface_plot} and \code{profile_plot} methods may be used to obtain and plot $\hat{f}(\bm{s},h,t)$ of Equation (\ref{eq:krig_exp}). In this case, the user has to provide the corresponding covariate (\code{X_beta}) for the scale/vector \code{h}, time \code{t} or location $\bm{s}$ (\code{lon0}, \code{lat0}) of interest.

\begin{figure}
\centering
\includegraphics[width=1\textwidth]{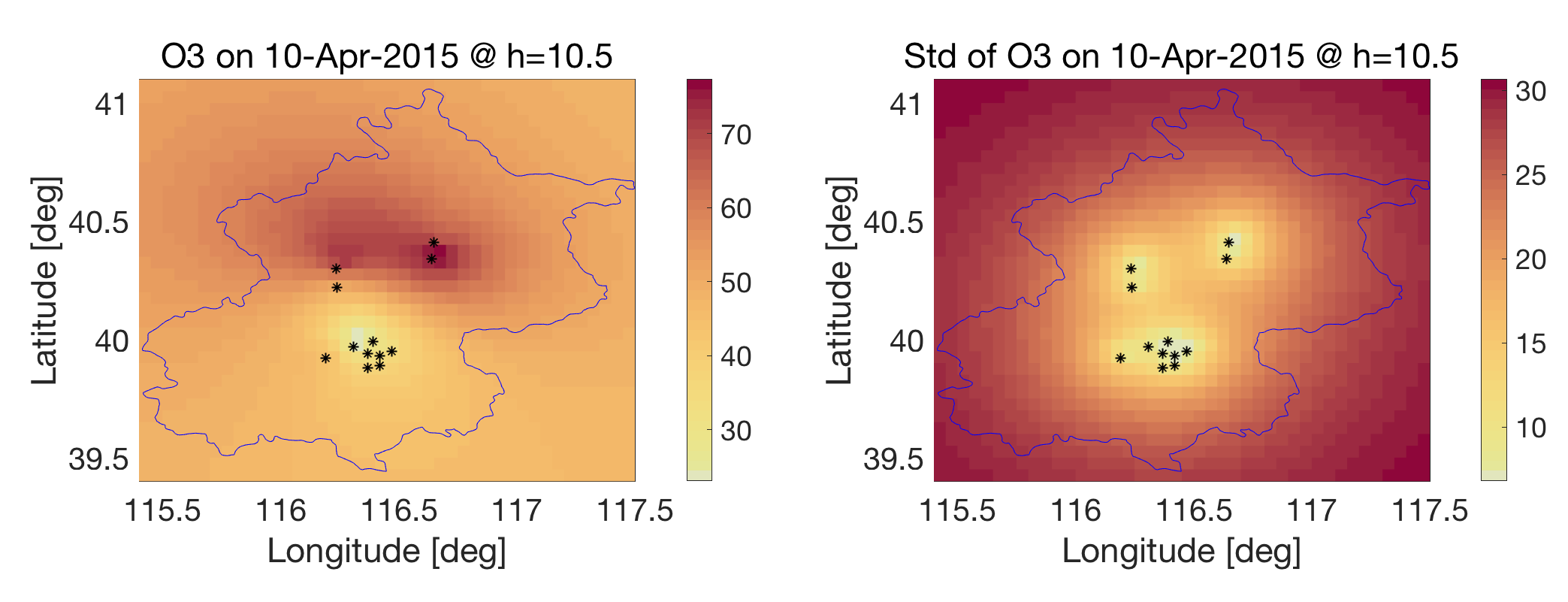}
\caption{$O_3$ concentrations and their standard deviation at 10:30 am ($h = 10.5$), on 10 April 2015, where 12 stations are marked with black stars.}
\label{fig:krig_surface_plot}
\end{figure}

Specifically, the \code{surface_plot} method is used to display the $O_{3}$ map using \code{h}, \code{t}, \code{X_beta} as input arguments. In the case of unavailable \code{X_beta}, the mapping concerns the component $\bm{\phi}(h)^{\top}\bm{z}(\bm{s},t)$. Loaded from the homonym file, the array \code{X_beta_t_100} refers to time $t=100$ and hour $h=10.5$ and has the dimension $35 \times 43 \times 3$. Maps of $O_{3}$ concentrations and their standard deviation are shown in Figure \ref{fig:krig_surface_plot}.

\begin{CodeChunk}
\begin{CodeInput}
load ../Data/kriging/X_beta_t_100;
t = 100; 
h = 10.5; 
[y_hat, diag_Var_y_hat] = o_krig_result.surface_plot(h, t, X_beta_t_100);
\end{CodeInput}
\end{CodeChunk}

\begin{figure}
\centering
\includegraphics[width=0.9\textwidth]{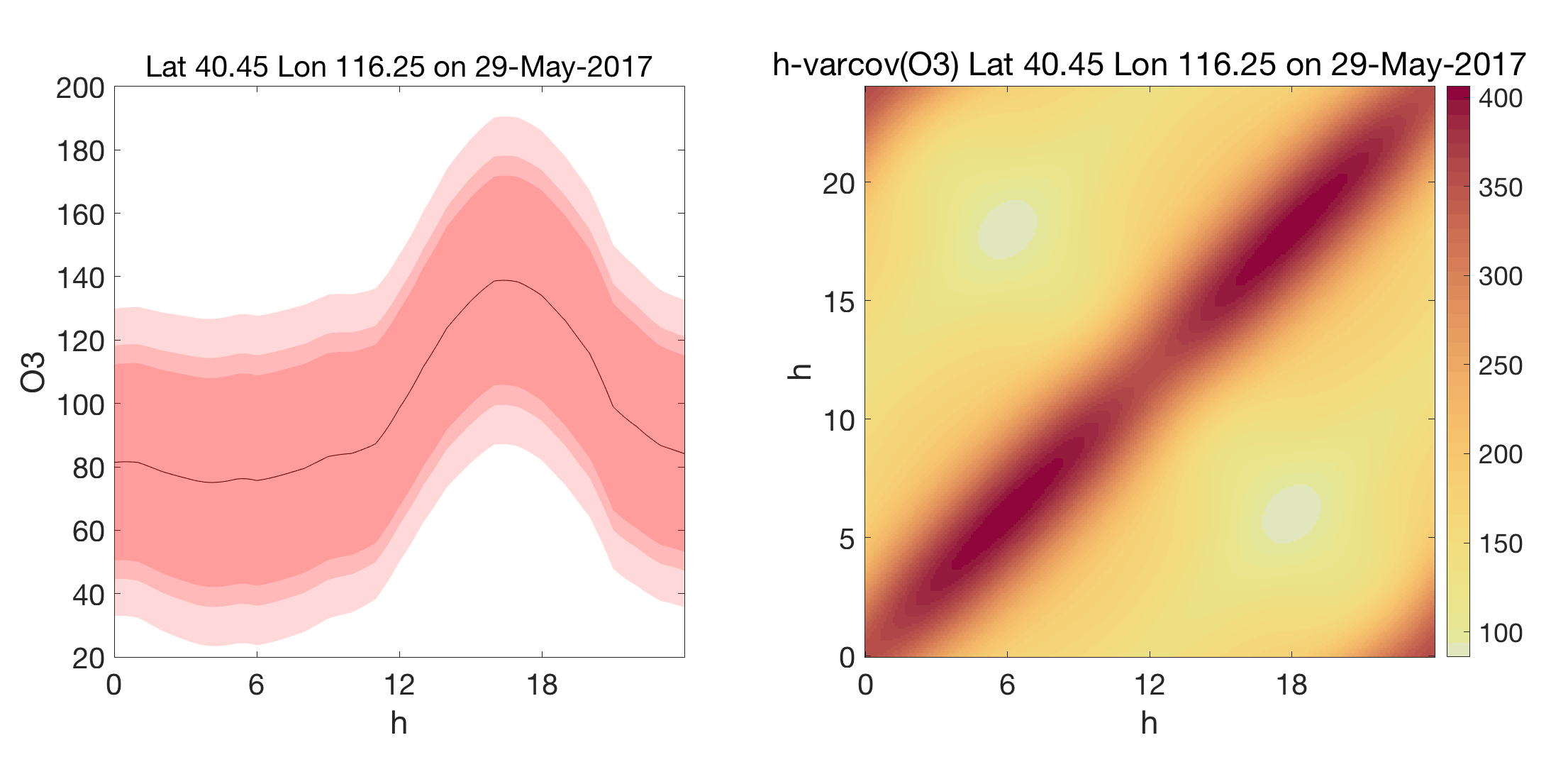}
\caption{$O3$ concentrations with $90\%, 95\%, 99\%$ confidence bands (different shadings), and their variance--covariance at latitude 40.45, longitude 116.25 on 29 May 2017.}
\label{fig:krig_profile_plot}
\end{figure}

On the other hand, the \code{profile\_plot} method is used to display the $O_{3}$ profile at a given spatial location $\bm{s}$ (\code{lon0}, \code{lat0}) and time \code{t}. Still, the profile plot concerns the component $\bm{\phi}(h)^{\top}\bm{z}(\bm{s},t)$ if \code{X\_beta} is not provided. After loading the \code{X\_beta\_h} (dimension $25\times 3$) from the homonym file, this method represents the profile of $O_{3}$ concentrations and their variance--covariance matrix as in Figure \ref{fig:krig_profile_plot}:

\begin{CodeChunk}
\begin{CodeInput}
load ../Data/kriging/X_beta_h;
h = 0:24;
lon0 = 116.25;
lat0 = 40.45;
t = 880;
[y_hat, diag_Var_y_hat] = o_krig_result.profile_plot(h, lon0, lat0, ...
	t, X_beta_h);
\end{CodeInput}
\end{CodeChunk}

Note that the prediction in Equation (\ref{eq:krig_exp}) and the variance in Equation (\ref{eq:krig_var}) are stored in the output arguments \code{y_hat}, and \code{diag_Var_y_hat}, respectively.

\section[Case study on climate data]{Case study on climate data}\label{sec:casestudy_climate}

In order to show the complexity-reduction capabilities of \pkg{D-STEM} v2, a data set of temperature vertical profiles collected by the radiosondes of the Universal Radiosonde Observation Program (RAOB) is now considered. 
The profiles are observed over the Earth's
sphere, and they are misaligned, that is, each profile differs in terms of the number
of observations and altitude above the ground of each observation.
Additionally, the computation burden is higher due to the higher number of
spatial locations at which profiles are observed.

\subsection{RAOB data}

Radiosondes are routinely launched from stations all over the world to
measure the state of the upper troposphere and lower stratosphere. Data collected 
by radio sounding have applications in weather prediction and climate studies.

Temperature data from 200 globally distributed stations collected daily
during January 2015 at 00:00 and 12:00 UTC are considered here. 
Each profile consists of a given number of measurements taken at different
pressure levels. Since the weather balloon carrying the radiosonde usually
explodes at an unpredictable altitude, the profile measurements
are misaligned across the profiles and have different pressure ranges. A
functional data approach is natural in this case since the underlying temperature
profile can be seen as a continuous function sampled at some pressure levels. 

Figure \ref{fig:RAOB_data} depicts the spatial locations of temperature measurements
taken on 1 January 2015 at 00:00 UTC. This demo data set, which only covers one month,
includes around $10^5$ data points. When the full data set is used in climate studies,
the number of data points grows to around $10^8$. In this case, a recent server machine with multiple CPUs
with at least 256 GB of RAM is required for model estimation and kriging.

The focus of the case study is on the difference between the radiosonde
measurement and the output of the ERA-Interim global atmospheric reanalysis
model provided by ECMWF. In particular, the aim is to study the spatial
structure of the this difference in 4D space, where the 
dimensions are latitude, longitude, altitude and time.

\begin{figure}
\centering
\includegraphics[width=0.75\textwidth]{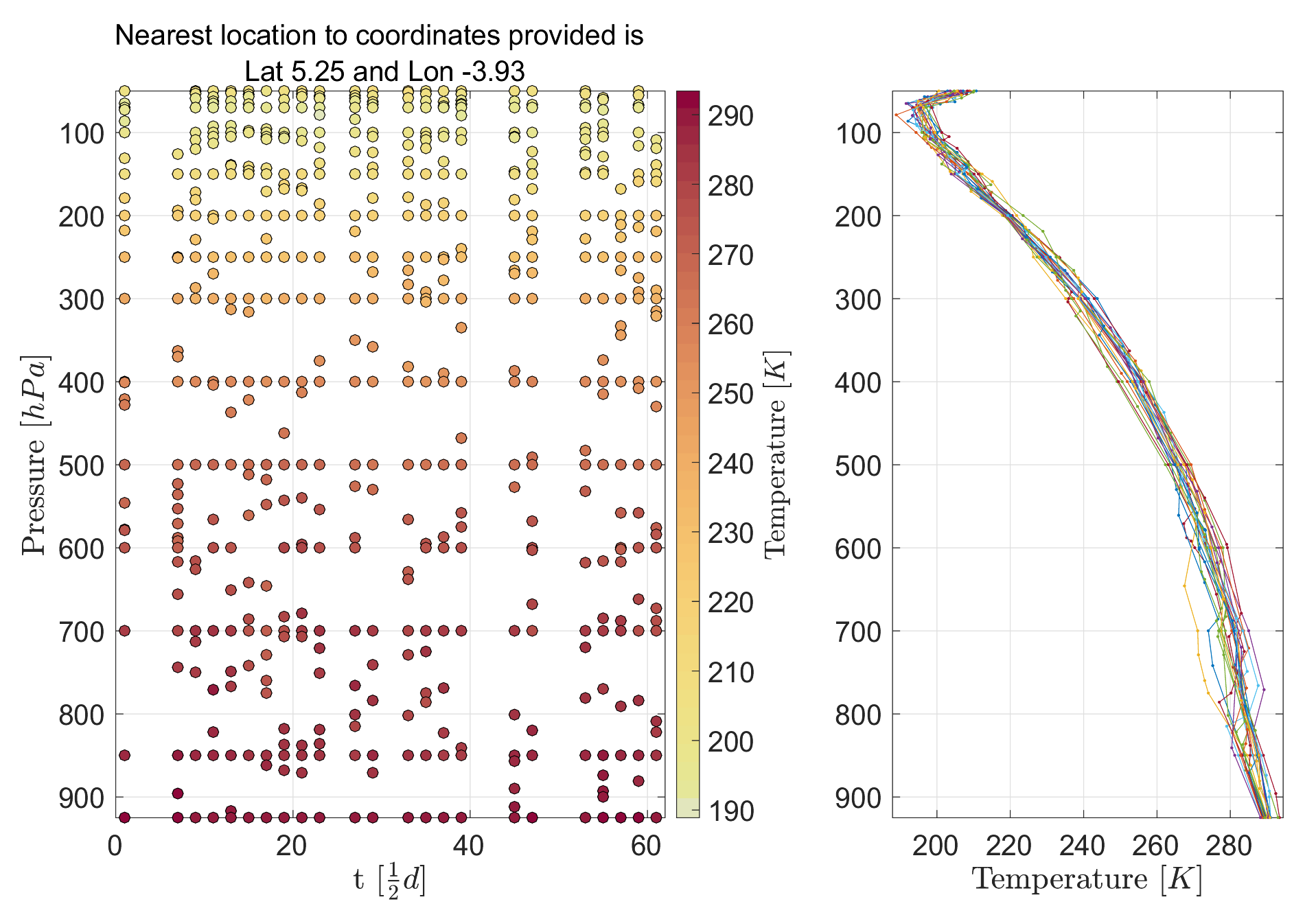}
\caption{\label{fig:data_profile_raob} Temperature at location 5.25 latitude and -3.93 longitude in January 2015. Left: each dot is a temperature measurement. The colour of the dot depicts the temperature. Right: each graph is a temperature vertical profile collected through radio sounding.}
\end{figure}

The model for temperature $y$ is as follows%
\[
y \left(  \bm{s},h,t\right)  =x_{ERA}\left(  \bm{s},h,t\right)
\beta_{ERA}\left(  h\right)  +\bm{\phi}\left(  h\right)  ^{\top}\bm{z}\left(  \bm{s},t\right)  +\varepsilon\left(  \bm{s}%
,h,t\right),
\]
where $h\in\left[ 50, 925\right]  $ $hPa$ is the pressure level, while
$t=1,...,62$ is a discrete time index for January 2015. Figure \ref{fig:data_profile_raob} shows the temperature measurements at a given station, where 50 and 925 $hPa$ correspond approximately to 25 and 1.3 km, respectively.

\subsection{Software implementation}

This section details the software implementation of the case study described above as in script \\
\code{demo_section5.m}, which can be also executed in the \code{demo_menu_user.m} script. To avoid repetition, only the relevant parts of the script that differ from the case study of Section \ref{sec:casestudy_ozone} are reported and commented on here. In particular, data
loading and instantiation of the \code{stem_model} object are not described.

\subsubsection{Model estimation}

The problem of vertical misalignment of the measurements is completely transparent to the user and is handled by the internal \code{stem_misc.data_formatter} method when creating the \code{stem_data} object.
Note that the dimension of the matrices in \code{o_varset} depends on $q$, the maximum number of measurements in each profile. 
To prevent out-of-memory problems, it is advisable to avoid data sets in which only a few profiles have a large number of
measurements, which could result in large matrices in
\code{o_varset}, with most of the elements set to \code{NaN}.

B-spline bases are used, since, in this application, vertical profiles are not periodic with respect to the pressure domain. The corresponding object of class \code{stem_fda} is created in the following way:%

\begin{CodeChunk}
\begin{CodeInput}
spline_order = 2;
rng_spline = [50,925];
knots_number = 5;
knots = linspace(rng_spline(1),rng_spline(2),knots_number);

input_fda.spline_type = 'Bspline';
\end{CodeInput}
\vspace{-.85cm} 
\begin{CodeInput}
input_fda.spline_order = spline_order;
input_fda.spline_knots = knots;
\end{CodeInput}
\vspace{-.85cm} 
\begin{CodeInput}
input_fda.spline_range = rng_spline;
o_fda = stem_fda(input_fda);
\end{CodeInput}
\end{CodeChunk}

Note that the knots are equally spaced along the functional range. In general,
however, non-equally spaced knots can be provided, and each model component
(i.e. $\sigma_{\varepsilon}^{2}$, $\beta_{j}$ and $\bm{\phi}(h)^{\top}\bm{z}(\bm{s}%
,t)$) can have a different set of knots. This is obtained using \code{spline_order} and \code{spline_knots} with additional suffixes \code{_sigma}, \code{_beta}, \code{_z}.

Although this data set is not large, the demo shows how to
enable the partitioning discussed in Section \ref{sec:partitioning}. First, the spatial locations are partitioned using the modified k-means algorithm:%

\begin{CodeChunk}
\begin{CodeInput}
k = 5;
trials = 100;
lambda = 5000;
partitions = o_data.kmeans_partitioning(k, trials, lambda);
\end{CodeInput}
\end{CodeChunk}

where \code{k} is the number of partitions, \code{trials} is the number of times when the k-means
algorithm is executed starting from randomised centroids and \code{lambda} is $\lambda$ in Equation (\ref{eq:k-means}).

At the end of the k-means algorithm, data are internally re-ordered for parallel
computing. Model estimation is done after creating and setting an object of class
\code{stem_EM_options}. To do this, the output of the \code{kmeans_globe} method is passed to the
\code{partitioning_block_size} property of the \code{o_EM_options} object.
Additionally, for parallel computing, the number of workers must be set to a
value higher than 1. In general, this could be any number up to the number of
cores available on the machine.

\begin{CodeChunk}
\begin{CodeInput}
o_EM_options = stem_EM_options();
o_EM_options.partitions = partitions;
o_EM_options.workers = 2;
o_model.EM_estimate(o_EM_options);
\end{CodeInput}
\end{CodeChunk}

\begin{figure}
\centering
\includegraphics[width=0.80\textwidth]{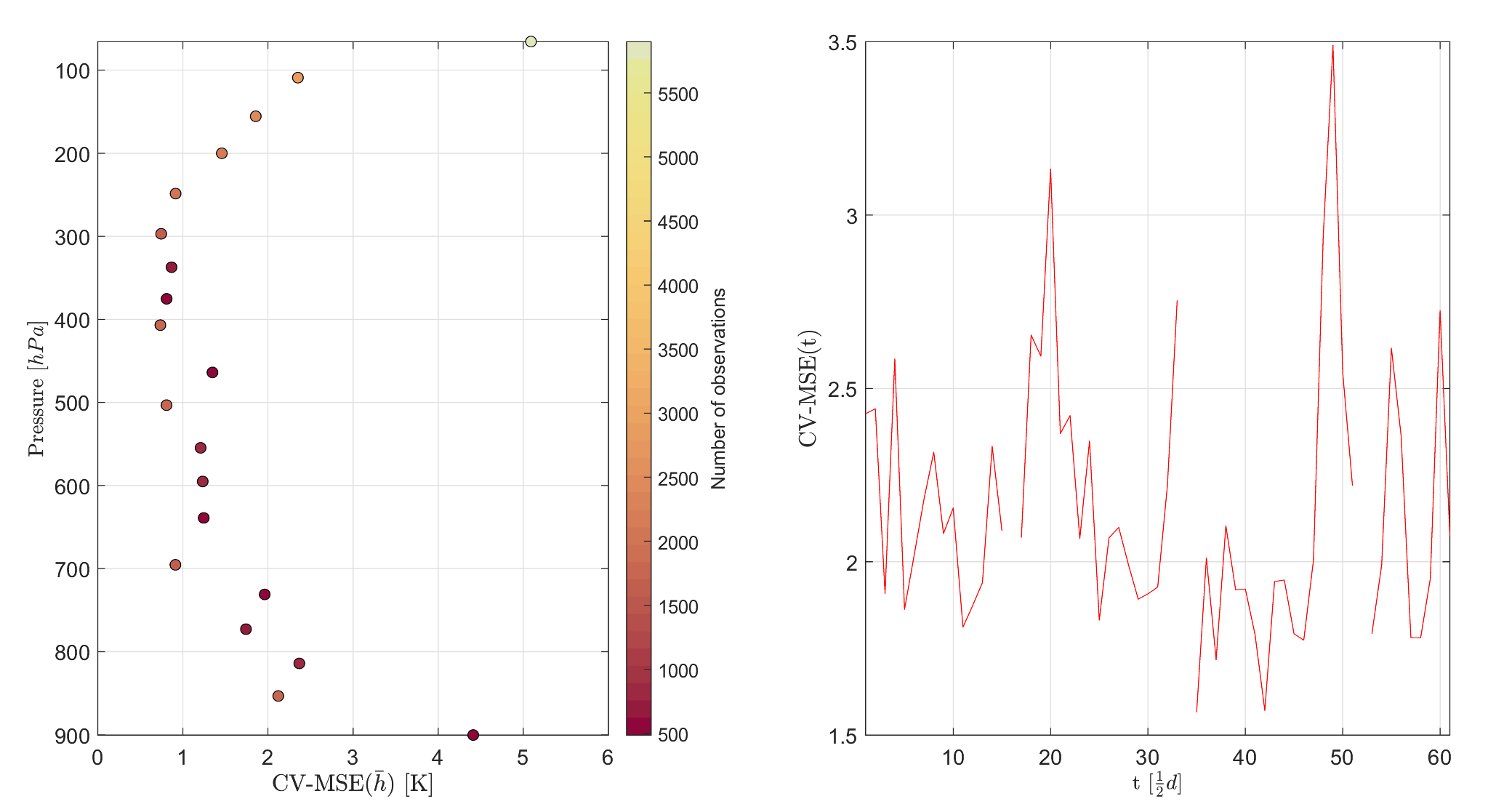} \caption{(Left) Validation MSE with respect to the $\bar{h}$ coloured by the number of observations $n_b$, and (Right) the validation MSE with respect to time $t$. }
\label{fig:MSE_raob_h_and_t}%
\end{figure}

\begin{figure}[ptb]
\centering
\includegraphics[width=0.95\textwidth]{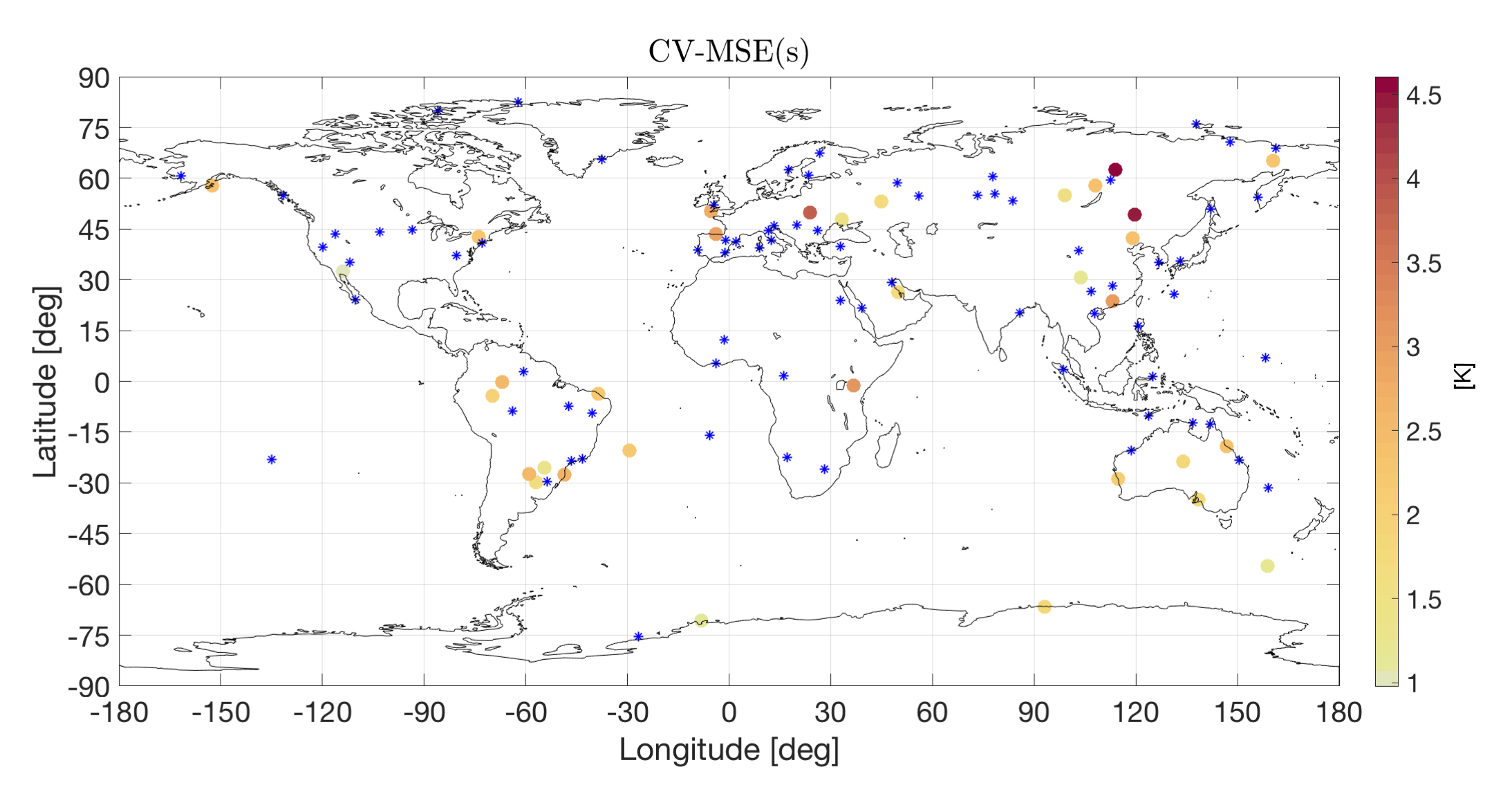} \caption{Validation MSE for the thirty-three stations, where the stations used for estimation are marked with blue stars.}
\label{fig:MSE_raob_s_scatter}%
\end{figure}

The three validation MSEs defined in Section \ref{sec:val} are shown in Figure \ref{fig:MSE_raob_h_and_t} and \ref{fig:MSE_raob_s_scatter}. To generate these figures, the method \code{plot_validation} is called with vertical = 1, which provides ``atmospheric profile'' plots with $h$ on the vertical axis:

\begin{CodeChunk}
\begin{CodeInput}
vertical = 1;
o_model.plot_validation(vertical);
\end{CodeInput}
\end{CodeChunk}

\subsubsection{Kriging}

Interpolation across space and over time is done as in Section
\ref{sec:O3_kriging}. However, complexity reduction is enabled by adopting the
nearest neighbour approach detailed in Section \ref{sec:kriging}.

To do this, a class constructor is first called, where the \code{block_size} is used to define the number of spatial locations in $\mathcal{S}_{l}^{\ast}$, and then \code{nn_size} is used to define $\tilde{n}$. Additionally, setting
\code{o_krig_options.workers} makes it possible to do the kriging over the $u$ blocks in
parallel using up to the allocated number of workers:

\begin{CodeChunk}
\begin{CodeInput}
o_krig_options = stem_krig_options();
o_krig_options.block_size = 150;
o_krig_options.nn_size = 10;
o_krig_options.workers = 2;
\end{CodeInput}
\end{CodeChunk}

Finally, kriging predictions and standard errors are mapped for a given $h\in\mathcal{H}$ and time $t$:

\begin{CodeChunk}
\begin{CodeInput}
h = 875.3; 
t = 12;
o_krig_result.surface_plot(h, t);
\end{CodeInput}
\end{CodeChunk}

\begin{figure}
\centering
\includegraphics[width=0.95\textwidth]{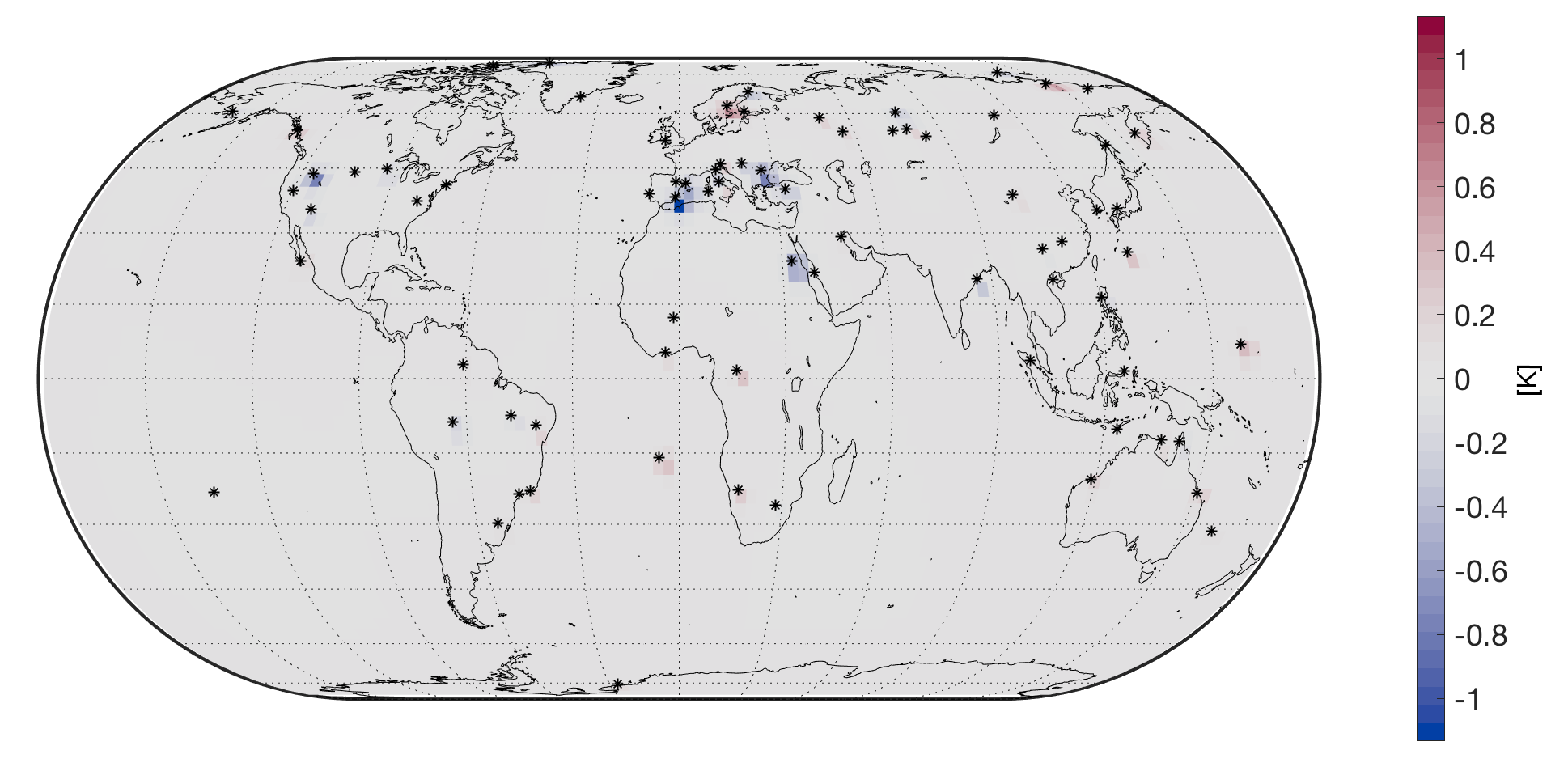}
\caption{\label{fig:krig_globe_mean} $\bm{\phi}(h)^{\top}\bm{\hat{z}}(\bm{s},t)$ at pressure $875.3$ $hPa$, and 12:00 am on 6 January 2015, where $200$ stations are shown as black stars.}
\end{figure}

\begin{figure}
\centering
\includegraphics[width=0.95\textwidth]{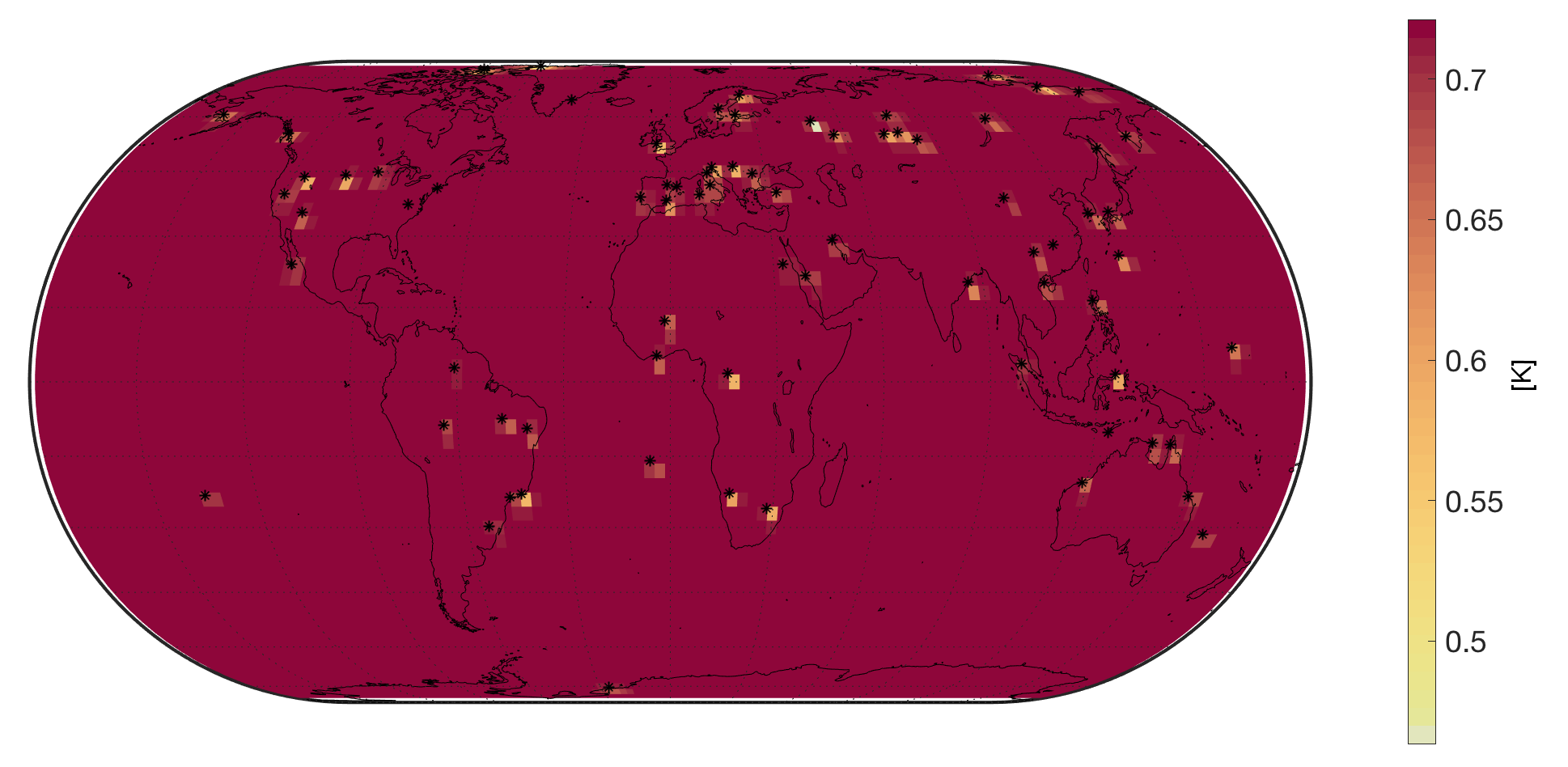}
\caption{\label{fig:krig_globe_std} Standard deviation of $\bm{\phi}(h)^{\top}\bm{\hat{z}}(\bm{s},t)$ at pressure $875.3$ $hPa$, and 12:00 am on 06 January 2015, where $200$ stations are shown as black stars.}
\end{figure}

Since covariates are not provided to the \code{surface_plot} method, the plots are on the component
$\bm{\phi}(h)^{\top}\bm{\hat{z}}(\bm{s},t)$, namely, the difference between RAOB and ERA-Interim and its standard deviation. The output
of the above code is depicted in Figures \ref{fig:krig_globe_mean} and \ref{fig:krig_globe_std}.

\section[Concluding remarks]{Concluding remarks}
\label{sec:remarks}

This paper introduced the package \pkg{D-STEM} v2 through two case studies of spatio-temporal modelling of functional data. It is shown that, in addition to maximum likelihood estimation, Hessian approximation and kriging for large
data sets, \pkg{D-STEM} v2 also develops several data-handling capabilities, allows for automatic construction of relevant objects and provides graphical output. In particular, it provides high-quality global maps and two kinds of functional
plotting: the traditional x--y plot and the vertical profile plot, which is popular, for example, in atmospheric data analysis. 
In this regard, model validation and kriging are straightforward.

\pkg{D-STEM} v2 fills a gap in functional geostatistics. In fact, although statistical methods for georeferenced functional data have been recently
developed (e.g. \cite{ignaccolo2014kriging}), standard geostatistical packages do not consider functional data, especially in the spatio-temporal context.

The successful use of \pkg{D-STEM} v1 in a number of applications proved that the EM algorithm implementation is quite stable. Now, due to improvements in computational efficiency, the new \pkg{D-STEM} v2 has the capability to handle large data sets.
Moreover, thanks to the approximated variance--covariance matrix, it is possible to compute standard errors for all model parameters relatively fast and avoid the large number of iterations typically required by an MCMC approach for making inferences. \\
However, a limit of the EM algorithm is its limited flexibility to changes in the model equations. Indeed, changes in parametrisation or latent variable structure usually require deriving new closed-form estimation formulas and changing the software accordingly. Moreover, changes in covariance functions are not easy to handle. \\
Computationally, the main limit of \pkg{D-STEM} v2 is in the number $p$ of basis functions that can be handled. 
Even if partitioning is exploited in $k$ blocks of size $r$, computational complexity is $\mathcal{O}\left(Tkr^{3}p^{3}\right)$, meaning that $p$ cannot be large.

Currently, the authors are working on a new version, which makes it possible to handle multivariate functional space-time data and user-defined spatial covariance functions, which will make \pkg{D-STEM} v2 a valid and comprehensive alternative to the Gaussian process regression models (\code{fitrgp}) implemented in the Statistics and Machine Learning Toolbox of \proglang{MATLAB}.

\section*{Acknowledgments}

\begin{leftbar}
We would like to thank China's National Key Research Special Program Grant (No. 2016YFC0207701) and the Center for Statistical Science at Peking University. 

\end{leftbar}


\bibliography{JSS_2020_R2}

\typeout{get arXiv to do 4 passes: Label(s) may have changed. Rerun}

\end{document}